\documentclass[12pt]{article}
\usepackage{amssymb}

\def\be{\begin{equation}}
\def\ee{\end{equation}}
\def\bea{\begin{eqnarray}}
\def\eea{\end{eqnarray}}
\def\ba{\begin{array}}
\def\ea{\end{array}}

\catcode`\@=11
\def\@citex[#1]#2{%
\if@filesw \immediate \write \@auxout {\string \citation {#2}}\fi
\@tempcntb\m@ne \let\@h@ld\relax \def\@citea{}%
\@cite{%
  \@for \@citeb:=#2\do {%
    \@ifundefined {b@\@citeb}%
      {\@h@ld\@citea\@tempcntb\m@ne{\bf ?}%
      \@warning {Citation `\@citeb ' on page \thepage \space undefined}}%
      {\@tempcnta\@tempcntb \advance\@tempcnta\@ne%
      \@tempcntb\number\csname b@\@citeb \endcsname \relax%
      \ifnum\@tempcnta=\@tempcntb %Number follows previous--hold on to it
        \ifx\@h@ld\relax%
          \edef \@h@ld{\@citea\csname b@\@citeb\endcsname}%
        \else%
          \edef\@h@ld{\ifmmode{-}\else--\fi\csname b@\@citeb\endcsname}%
        \fi%
      \else%   %  non-successor--dump what's held and do this one
        \@h@ld\@citea\csname b@\@citeb \endcsname%
        \let\@h@ld\relax%
      \fi}%
    \def\@citea{,\penalty\@highpenalty\,}%
  }\@h@ld
}{#1}}

\def\@citeb#1#2{{[#1]\if@tempswa , #2\fi}}
\def\@citeu#1#2{{$^{#1}$\if@tempswa , #2\fi }}
\def\@citep#1#2{{#1\if@tempswa , #2\fi}}

\def\bcites{         % cite with []'s
        \catcode`\@=11
        \let\@cite=\@citeb
        \catcode`\@=12
}

\def\upcites{         % cite with exponents
        \catcode`\@=11
        \let\@cite=\@citeu
        \catcode`\@=12
}

\def\plaincites{      % cite without brackets
        \catcode`\@=11
        \let\@cite=\@citep
        \catcode`\@=12
}

\newcount\hour
\newcount\minute
\newtoks\amorpm
\hour=\time\divide\hour by 60
\minute=\time{\multiply\hour by 60 \global\advance\minute by-\hour}
\edef\standardtime{{\ifnum\hour<12 \global\amorpm={am}%
        \else\global\amorpm={pm}\advance\hour by-12 \fi
        \ifnum\hour=0 \hour=12 \fi
        \number\hour:\ifnum\minute<10 0\fi\number\minute\the\amorpm}}
\edef\militarytime{\number\hour:\ifnum\minute<10 0\fi\number\minute}

\def\draftlabel#1{{\@bsphack\if@filesw {\let\thepage\relax
   \xdef\@gtempa{\write\@auxout{\string
      \newlabel{#1}{{\@currentlabel}{\thepage}}}}}\@gtempa
   \if@nobreak \ifvmode\nobreak\fi\fi\fi\@esphack}
        \gdef\@eqnlabel{#1}}
\def\@eqnlabel{}
\def\@vacuum{}
\def\marginnote#1{}
\def\draftmarginnote#1{\marginpar{\raggedright\scriptsize\tt#1}}
\def\draft{
        \pagestyle{plain}
        \overfullrule=2pt
        \oddsidemargin -.5truein
        \def\@oddhead{\sl \phantom{\today\quad\militarytime} \hfil
        \smash{\Large\sl DRAFT} \hfil \today\quad\militarytime}
        \let\@evenhead\@oddhead
        \let\label=\draftlabel
        \let\marginnote=\draftmarginnote
        \def\ps@empty{\let\@mkboth\@gobbletwo
        \def\@oddfoot{\hfil \smash{\Large\sl DRAFT} \hfil}
        \let\@evenfoot\@oddhead}
        \def\@eqnnum{(\theequation)\rlap{\kern\marginparsep\tt\@eqnlabel}%
        \global\let\@eqnlabel\@vacuum}  }
\usepackage{mathptmx}       % selects Times Roman as basic font
\usepackage{helvet}         % selects Helvetica as sans-serif font
\usepackage{courier}        % selects Courier as typewriter font
\usepackage{type1cm}        % activate if the above 3 fonts are 
\usepackage{makeidx}         % allows index generation
\usepackage{graphicx}        % standard LaTeX graphics tool
\usepackage{multicol}        % used for the two-column index
\usepackage[bottom]{footmisc}% places footnotes at page bottom

\usepackage{amsmath,amssymb}
\usepackage{bm}
\makeindex             % used for the subject index
\def\siohalf{\frac{1}{2}}

\def\sio4{\frac{1}{4}}

\def\siovereq#1#2{\lower3pt\vbox{\baselineskip1.5pt \lineskip1.5pt
                                                                                
\ialign{$#1\hfill##\hfil$\crcr#2\crcr\sim\crcr}}}

\def\siobea{\begin{eqnarray}}
\def\sioeea{\end{eqnarray}}
\def\siobeq{\begin{eqnarray}}
\def\sioeeq{\end{eqnarray}}
\def\siobe{\begin{equation}}
\def\sioee{\end{equation}}

\def\sioba{\begin{array}}
\def\sioea{\end{array}}
\def\siobi{\begin{itemize}}
\def\sioei{\end{itemize}}

\title{Analytic calculation of quasi-normal modes\footnote{Research supported in part by the US Department of Energy under grant DE-FG05-91ER40627.}}
\author{George Siopsis\footnote{E-mail: siopsis@tennessee.edu}\\
\em Department of Physics
and Astronomy, \\
\em The University of Tennessee, Knoxville, \\
\em TN 37996 - 1200, USA.
}
\date{}
\begin{document}
                                                                                
\maketitle
\vspace{-3.5in}\hfill UTHET-08-0201\vspace{3.5in}
                                                                                
\abstract{We discuss the analytic calculation of quasi-normal modes of various types of perturbations of black holes both in asymptotically flat and anti-de Sitter spaces.
We obtain asymptotic expressions and also show how corrections can be calculated perturbatively. We pay special attention to low-frequency modes in anti-de Sitter space because they govern the hydrodynamic properties of a gauge theory fluid according to the AdS/CFT correspondence. The latter may have experimental consequencies for the quark-gluon plasma formed in heavy ion collisions.}

\vspace{2in}

\noindent{\sl Prepared for the proceedings of the 4th Aegean Summer School on Black Holes, Mytilene,
Greece, September 2007.}
\newpage

\section{Introduction}
\label{siosec:1}

\begin{quotation}
To many practitioners of quantum gravity the black hole plays the role of a
soliton, a non-perturbative field configuration that is added to the spectrum
of particle-like objects only after the basic equations of their theory have
been put down, much like what is done in gauge theories of elementary particles,
where Yang-Mills equations with small coupling constants determine the
small-distance structure, and solitons and instantons govern the large-distance
behavior.

Such an attitude however is probably not correct in quantum gravity. The coupling
constant increases with decreasing distance scale which implies that the
smaller the distance scale, the stronger the influences of ``solitons''.
At the Planck scale it may well be impossible to disentangle black holes from
elementary particles.

\rightline{-- G.~'t Hooft}
\end{quotation}
Quasi-normal modes (QNMs) describe small perturbations of a black hole which
is a thermodynamical system whose (Hawking) temperature and
entropy are given in terms of its global characteristics (total mass, charge
and angular momentum).
They are obtained by solving
a wave equation for small fluctuations subject to the conditions that the
flux be 
ingoing at the horizon and 
outgoing at asymptotic infinity.
These boundary conditions in general lead to a
discrete spectrum of complex frequencies whose
imaginary part determines the decay time of the small fluctuations
\siobe \Im \omega = \frac{1}{\tau}\sioee
There is a vast literature on quasi-normal modes and we make no attempt to review it. Instead, we concentrate on obtaining analytic expressions for quasi-normal modes of various black hole perturbations of interest. One can rarely obtain analytic expressions in closed form. Instead, we discuss techniques which allow one to calculate the spectrum perturbatively starting with an asymptotic regime (e.g., high or low overtones). In asymptotically flat space, we discuss the cases of four-dimensional Schwarzschild and Kerr black holes. Generalization to higher-dimensional spacetimes does not present substantially new calculational challenges. However, we should point out that the case of a rotating black hole is considerably harder than the Schwarzschild case.

We also discuss asymptotically AdS spaces and obtain the spectrum as a perturbative expansion around high overtones. At leading order the frequencies are proportional to the radius of the horizon. When expanding around low overtones, one in general obtains an additional frequency which is inversely proportional to the horizon radius. Thus for large black holes there is a gap between the lowest frequency and the rest of the spectrum of quasi-normal modes.
We pay special attention to the lowest frequencies because they govern the behavior of the gauge theory fluid on the boundary per the AdS/CFT correspondence.
The latter may have experimental consequencies pertaining to the formation of the quark-gluon plasma in heavy ion collisions.

\section{Flat spacetime}
\label{siosec:2}

%
%\section{Flat spacetime}
%\label{siosec:2}
We start with a study of QNMs in asymptotically flat space-times.
We discuss scalar perturbations of Schwarzschild and Kerr black holes in four dimensions.

\subsection{Schwarzschild black holes}

%\section{Introduction}

The metric of a Schwarzschild black hole in four dimensions is
\siobe
ds^2 = -f(r)\, dt^2 + \frac{dr^2}{ f(r)} +r^2 d \Omega^2
\ \ , \ \
f(r) = 1 - \frac{2G M}{r}\sioee
%where $G$ is Newton's constant and $M$ is the mass of the black hole.
The Hawking temperature is
\siobe
T_H = \frac{1}{8\pi GM} = \frac{1}{4\pi r_0}
\sioee
where $r_0 = 2GM$ is the radius of the horizon.

A spin-$j$ perturbation of frequency $\omega$ is governed by the radial equation
\siobe
-f(r) \frac{d}{d r} \left( f(r)\frac{d\Psi}{d r}
\right) + V(r) \Psi  =\omega^2 \Psi \sioee
where $V(r)$ is the ``Regge-Wheeler'' potential
\siobe V(r) = f(r) \left( \frac{\ell (\ell+1)}{r^2} + \frac{(1-j^2)r_0}{r^3} \right)\sioee
The spin is $j=0,1,2$ for scalar, electromagnetic and gravitational perturbations, respectively.
%and gravitational wave, respectively. We shall keep the discussion general.
%In fact, it will be necessary to 
It is advantageous to avoid integer values of $j$ throughout
the discussion and only take the limit
$j\to $ integer
at the end of the
calculation.
%\newpage

By defining the ``tortoise coordinate''
\siobe r_* = \int \frac{dr}{f(r)} = r + r_0 \ln \left(\frac{r}{r_0} -1 \right) \sioee
the wave equation may be brought into a Schr\"odinger-like form,
\siobe
-\frac{d^2\Psi}{d r_*^2} + V(r(r_*)) \Psi  =\omega^2 \Psi \sioee
to be solved along the entire real $r_*$-axis.
At both ends the potential vanishes
($V\to 0$ as $r_*\to \pm\infty$)
therefore the solutions behave as
$\Psi\sim e^{\pm i\omega r_*}$.
For QNMs, we demand
\siobe \Psi \sim e^{\mp i\omega r_*} \ \ , \ \ r_* \to \pm\infty\sioee
assuming $\Re \omega > 0$.

\subsubsection{Limit $\ell\to \infty$}

In this case it suffices to consider the potential near its maximum.
Expanding around the
maximum of the potential ($V_0' (r_{max}) = 0$) \cite{siob-ferrari-mashhoon},
\siobe r_{max} = \frac{3}{2}\, r_0 +
\mathcal{O} (1/\ell) \,,\sioee
we obtain
\siobe V_0 [ r(r_*)] \approx \alpha^2 - \beta^2
(r_*-r_*(r_{max}))^2 \,,\sioee where
\siobe \alpha^2 =
\frac{4}{27} \left( \ell + \siohalf \right) r_0^2 + \mathcal{O} (1/\ell )
\ \ , \ \ \ \
\beta^2 = \frac{16}{729} \left( \ell + \siohalf
\right) + \mathcal{O} (1/\ell )\,.\sioee
The solutions to the wave equation are
\siobe \Psi_n = H_n(\sqrt{i\beta} x)
e^{i\beta x^2/2} \ \ , \ \ n = 0,1,2,\dots\sioee where $H_n$ are
Hermite polynomials.
The
corresponding eigenvalues are
\siobe \omega_n =\frac{2}{3\sqrt{3}\, r_0}
\left\{ \ell +\siohalf + i (n+\siohalf) \right\} + \mathcal{O} (1/\ell) \sioee
This result is in agreement with the standard WKB
approach~\cite{siob-konoplya}.
%\newpage
%\bi
%\item asymptotic form of QNM
%frequencies is related to the Barbero-Immirzi parameter of Loop Quantum Gravity.
%\ei
\subsubsection{Limit $n \to \infty$}

The asymptotic form of QNMs for large $n$ is
\siobe\label{sioe-1} \frac{\omega_n}{T_H} = (2n+1)\pi i + \ln 3 \sioee
independent of the angular momentum quantum number $\ell$.
This form was first
derived numerically
\cite{siob-Chandra-Det,siob-Leaver,siob-Nollert,siob-Andersson,siob-Bachelot}
%~\cite{bibn1,bibn2,bibn3,bibn4,bibn5}
and subsequently confirmed analytically
\cite{siob-Motl-Nei}.
The large imaginary part of the frequency
($\Im\omega_n$)
makes the numerical analysis cumbersome
but is easy to understand because the spacing of frequencies is $2\pi i T_H$
which is the same as the spacing of poles of a thermal Green function on the
Schwarzschild black hole background.
%\newpage
On the other hand the real part ($\Re\omega_n$) is small.
Its
analytical value was first proposed by Hod \cite{siob-Hod}.

%\newpage
The analytical derivation of the asymptotic form (\ref{sioe-1}) of QNMs by Motl and Neitzke \cite{siob-Motl-Nei} offered a new surprise
because it heavily
relied on the black hole singularity.
It is intriguing that the unobservable
region beyond the horizon influences the behavior of physical quantities.
%\newpage

We shall
calculate the asymptotic formula for QNMs including first-order
corrections \cite{siob-MS}
by solving the wave equation
perturbatively for arbitrary spin of the wave.
We shall obtain agreement with results from
numerical analysis for gravitational and scalar
waves \cite{siob-Nollert,siob-Ber-Kok}
and WKB analysis for gravitational waves
\cite{siob-vdB}.

Let
\siobe \Psi = e^{-i\omega r_*} f(r_*) ~. \sioee
We have
$f(r_*)\sim 1$ as $r_*\to +\infty$
and near the horizon,
$f(r_*)\sim e^{2i\omega r_*}$ (as $r_*\to -\infty$).
%\newpage
Let us continue $r$ analytically into
the complex plane and define the boundary condition at the horizon
in terms of the monodromy
of $f(r_*(r))$ around the singular point $r=r_0$,
\siobe \mathcal{M} (r_0) = e^{-4\pi\omega r_0}\sioee
along a contour running counterclockwise.
% This will serve as the definition of the boundary condition at the horizon.
%\newpage
We may deform the contour in the
complex $r$-plane so that it either lies
beyond the horizon ($\mathrm{Re} r
< r_0$) or
at infinity ($r\to \infty$).
The monodromy only gets a contribution from the segment lying beyond the
horizon.
%\newpage

It is convenient to change variables to
\siobe z = \omega (r_* -i\pi r_0) = \omega (r + r_0 \ln (1-r/r_0))\sioee
(where we chose a branch such that~$z\to 0$ as $r\to 0$).
The potential can be written as a series in $\sqrt{z}$,
\siobe V(z) = -\frac{\omega^2}{4z^2} \left( 1-j^2 + \frac{3\ell (\ell+1) +1- j^2
}{3}\, \sqrt{-\frac{2z}{\omega r_0}} + \dots \right)\sioee
which is a formal expansion in $1/\sqrt\omega$.

Now deform the
contour defining the monodromy so that it gets mapped onto the real axis in the $z$-plane.
Near the
singularity $z=0$,
\siobe z \approx -\frac{\omega}{2r_0}\, r^2\sioee
Choose a contour in the $r$-plane so that near $r=0$,
the positive and negative real axes in the $z$-plane are
mapped onto
\siobe \arg r = \pi - \frac{\arg\omega}{2} \ \ , \ \ \arg r = \frac{3\pi}{2}
- \frac{\arg\omega}{2} \sioee
in the $r$-plane, respectively.
These segments form a $\pi /2$ angle (independent of $\arg\omega$).

To avoid the $r=0$ singularity,
go around an arc of angle $3\pi /2$
which corresponds to an
angle of
$3\pi$ around $z=0$ in the $z$-plane.
%\newpage

Considering the black hole singularity ($r=0$), we note that there are two
solutions,
\siobe f_\pm (r) = r^{1\pm j} Z_\pm (r) \sioee
where $Z_\pm$ are analytic functions of $r$.
Going around an arc of angle of $3\pi/2$, we obtain
\siobe f_\pm (e^{3\pi i/2} r) = e^{3\pi (1\pm j) i/2} \, f_\pm (r)\sioee
which is an {\em exact} result.

To proceed further, we need to relate
the behavior of the wavefunction near the black hole singularity
to
its behavior at large $r$ in the complex $r$-plane.
To this end, we shall solve the wave equation perturbatively, thus writing
the wavefunction as a perturbation series in
$1/\sqrt\omega$.

%\newpage
At zeroth order, the wave equation reads
\siobe
\frac{d^2\Psi^{(0)}}{d z^2} + \left( \frac{1-j^2}{4z^2} + 1 \right) \Psi^{(0)}  = 0 \sioee
Two linearly independent solutions are
\siobe f_\pm^{(0)} (z) = e^{iz}\, \Psi_\pm^{(0)} = e^{iz} \sqrt{\frac{\pi z}{2}} J_{\pm j/2} (z)\sioee
in terms of Bessel functions.
We deduce the behavior at infinity ($z\to\infty$)
\siobe f_\pm^{(0)} (z) \sim e^{iz} \cos (z - \pi (1\pm j)/4)\sioee
%\newpage
The boundary conditions imply
$f(z)\sim$~const.~as $z\to\infty$
along the positive real axis in the $z$-plane.
Therefore, we ought to adopt the linear combination
\siobe f^{(0)} = f_+^{(0)} - e^{-\pi ji/2}\, f_-^{(0)} \sim e^{iz} \sqrt z\,
H_{j/2}^{(1)} (z)\sioee
(in terms of a Hankel function).
%which approaches a constant at infinity,
As $z\to \infty$, we obtain
\siobe f^{(0)}(z) \sim - e^{-\pi (1+j) i/4} \sin (\pi j/2) \sioee
a constant, as desired.

Going along the $3\pi $ arc around $z=0$ in the $z$-plane, we have
\siobe f^{(0)}(e^{3\pi i} z) = e^{3\pi (1+j)i/2} \left( f_+^{(0)} (z) - e^{-7\pi ji/2} \, f_-^{(0)} (z)\right) \sioee
As $z\to\infty$,
%whose behavior at infinity is found to be
\siobe f^{(0)}(z) \sim e^{-\pi (1+j) i/4} \sin (3\pi j/2)
+ e^{\pi (1-j)i/4} \sin (2\pi j) e^{2iz} \sioee
%using eq.~(\ref{eqff0}) again.
The monodromy to zeroth order is
\siobe \mathcal{M} (r_0) = -\frac{\sin (3\pi j/2)}{\sin (\pi j/2)}
= - (1+ 2\cos (\pi j)) \sioee
leading to a discrete set of complex frequencies (QNMs) \cite{siob-Motl-Nei}
\siobe\label{sioe-2} \frac{\omega_n}{T_H} = (2n+1)\pi i + \ln (1+2\cos(\pi j) ) + \mathcal{O} (1/\sqrt n) \sioee
%\newpage
Next, we calculate the first-order correction to the above expression \cite{siob-MS}.
Expanding the wavefunction in $1/\sqrt\omega$,
\siobe \Psi = \Psi^{(0)} + \frac{1}{\sqrt{-\omega r_0}}\, \Psi^{(1)} + \mathcal{O} (1/\omega)\sioee
the first-order correction obeys
\siobe
\frac{d^2\Psi^{(1)}}{d z^2} + \left( \frac{1-j^2}{4z^2} + 1 \right) \Psi^{(1)}  =\sqrt{-\omega r_0}\, \delta V \Psi^{(0)}\sioee
where
\siobe
\delta V(z) = \frac{1-j^2}{4z^2} + \frac{1}{\omega^2}\, V[r(z)] \sioee
Two linearly independent solutions are
\siobe \Psi_\pm^{(1)} (z) = \mathcal{C} \Psi_+^{(0)} (z)\int_0^z \Psi_-^{(0)}
\delta V \Psi_\pm^{(0)}
- \mathcal{C} \Psi_-^{(0)} (z)\int_0^z \Psi_+^{(0)}
\delta V \Psi_\pm^{(0)}\sioee
where
$\mathcal{C} = \frac{\sqrt{-\omega r_0}}{\sin (\pi j/2)}$
and the integral is along the positive real axis on the $z$-plane ($z>0$).
%We are assuming $0<j<1/2$ aiming at taking the limit $j\to 0^+$ eventually.
We obtain the large-$z$ behavior
\siobe \Psi_\pm^{(1)} (z) \sim c_{-\pm} \cos (z-\pi (1+ j)/4)
- c_{+\pm} \cos (z-\pi (1-j)/4)\sioee
where
\siobe c_{\pm\pm} = \mathcal{C} \int_0^\infty \Psi_\pm^{(0)} \delta V \Psi_\pm^{(0)}
\sioee
To obtain the small-$z$ behavior, expand
\siobe \delta V(z) = -\frac{3\ell (\ell+1) + 1-j^2 }{6\sqrt{-2\omega r_0} }\, z^{-3/2} + \mathcal{O} (1/\omega) \sioee
It follows that
\siobe \Psi_\pm^{(1)} = z^{1\pm j/2} G_\pm (z) + \mathcal{O} (1/\omega) \sioee
where $G_\pm$ are even analytic functions of $z$.

For the desired behavior as $z\to\infty$, define
\siobe \Psi = \Psi_+^{(0)} + \frac{1}{\sqrt{-\omega r_0}}\, \left\{ \Psi_+^{(1)} - e^{-\pi ji/2} \Psi_-^{(1)}
+ e^{-\pi ji/2} \xi\Psi_-^{(0)}\right\}
+ \dots \sioee
where $\xi  \sim \mathcal{O} (1)$
and dots represent terms of order higher than $\mathcal{O} (1/\sqrt\omega)$.
By demanding
$\Psi\sim e^{-iz}$ as $z\to +\infty$,
we fix
\siobe \xi =  \xi_+ + \xi_- \ \ , \ \
\xi_+ = c_{++} e^{\pi ji/2} - c_{+-}\ \ , \ \ 
\xi_- = c_{--} e^{-\pi ji/2} - c_{+-}\sioee
Then the requirement $f(z) = e^{iz} \Psi(z) \sim$ const.~as $z\to\infty$ yields
\siobe f(z) \sim - e^{-\pi (1+j)i/4} \sin (\pi j/2) \left\{ 1 - \frac{\xi_-}{\sqrt{-\omega r_0}} \right\} \sioee
%as $z\to +\infty$ along the real axis.
%\newpage
In the neighborhood of the black hole singularity
(around $z=0$), going around a $3\pi$ arc, we obtain
\siobe \Psi_\pm^{(1)} (e^{3\pi i} z) = e^{3\pi (2\pm j) i/2 } \Psi_\pm^{(1)} (z) \sioee
therefore
\siobea \Psi (e^{3\pi i} z) &=& \Psi^{(0)} (e^{3\pi i} z)
\nonumber\\
& &- e^{3\pi ji/2}
\frac{1}{\sqrt{-\omega r_0}}\,
\left\{ \Psi_+^{(1)} (z)
- e^{-7\pi ji/2} (\Psi_-^{(1)} (z) - i\xi \Psi_-^{(0)}
(z)) \right\} \sioeea
%\newpage
As $z\to \infty$ along the real axis,
\siobea f (z) &\sim & e^{-\pi (1+j)i/4} \sin (3\pi j/2) \left\{ 1-\frac{1}{\sqrt{-\omega r_0}}\, A\right\} \nonumber\\
& & + e^{\pi (1-j)i/4}
\sin(2\pi j) \left\{ 1-\frac{1}{\sqrt{-\omega r_0}}\, B\right\} e^{2iz}\nonumber\sioeea
where
%\be A = -c_{+-} (e^{3\pi ji/2}+e^{-3\pi ji/2} (1-2i)) + c_{++} e^{\pi ji/2} (e^{3\pi ji/2}- ie^{-3\pi ji/2})
%+ e^{-2\pi ji} c_{--} (1-i)\ee
%\be B = -c_{+-} e^{\pi ji/4} (1+e^{-4\pi ji} (1-2i)) + c_{++} e^{-\pi ji/4} (1-ie^{-3\pi ji})
%+ e^{-13\pi ji/4} c_{--} (1-ie^{\pi ji})\ee
\siobe A = \frac{i-1}{2}\ e^{\pi ji/2}\ \left( \xi_++i\xi_- -  \xi \, \cot (3\pi j/2)\right)\sioee
and $B$ is not needed for our purposes.
The monodromy to this order reads
\siobe \mathcal{M} (r_0) = - \frac{\sin (3\pi j/2)}{\sin (\pi j/2)}
\left\{ 1 + \frac{i-1}{2\sqrt{-\omega r_0}}\ e^{\pi ji/2}\ \left( \xi_- -\xi_+
+ \xi \cot(3\pi j/2) \right) \right\} \sioee
leading to the QNM frequencies \cite{siob-MS}
\siobea \frac{\omega_n}{T_H} &=& (2n+1)\pi i + \ln (1+2\cos(\pi j) )
+ \frac{e^{\pi ji/2}}{\sqrt{n+1/2}} \left( \xi_- -\xi_+
+ \xi \cot(3\pi j/2) \right) \nonumber\\
& & + \mathcal{O} (1/n) \sioeea
which
includes the $\mathcal{O} (1/\sqrt n)$ correction to the $\mathcal{O} (1)$ asymptotic expression (\ref{sioe-2}).
%\newpage

For an explicit expression, use
\siobe \mathcal{J} (\nu,\mu) \equiv \int_0^\infty dz\, z^{-1/2} J_\nu (z) J_\mu (z) = \frac{\sqrt{\pi/2}
\, \Gamma (\frac{\nu+\mu + 1/2}{2})}{\Gamma (\frac{-\nu+\mu+3/2}{2})
\Gamma (\frac{\nu+\mu+3/2}{2}) \Gamma (\frac{\nu-\mu +3/2}{2} )}\sioee
We obtain
\siobe c_{\pm\pm} = \pi\, \frac{3\ell (\ell + 1) +1-j^2}{12\sqrt 2\, \sin (\pi j/2)}
\, \mathcal{J} (\pm j/2, \pm j/2)\sioee
therefore
\siobea  \xi_- -\xi_+
+ \xi \cot(3\pi j/2)
 &=& (1-i) \ \frac{3\ell(\ell +1) +1-j^2}{24\sqrt 2 \pi^{3/2}}
\ \frac{\sin (2\pi j)}{\sin (3\pi j/2)}
 \nonumber\\
& & \times \Gamma^2 (1/4)\ \Gamma(1/4 + j/2)
\ \Gamma (1/4 - j/2)\sioeea
where we also used the identity 
$\Gamma (y) \Gamma (1-y) = \frac{\pi}{\sin (\pi y)}$.
%$\psi(1-z) - \psi (z) = \pi \cot \pi z$.
This expression has a well-defined finite limit as $j\to$ integer.
%\newpage

For scalar waves, let
$j\to 0^+$. We obtain
\siobe \frac{\omega_n}{T_H} = (2n+1)\pi i + \ln 3
 + \frac{1-i}{\sqrt{n+1/2}} \ \frac{\ell(\ell +1) +1/3}{6\sqrt 2 \pi^{3/2}}
\ \Gamma^4 (1/4) + \mathcal{O} (1/n) \sioee
which is in agreement with numerical results \cite{siob-Ber-Kok}.

For gravitational waves, we
let
$j\to 2$ and obtain
\siobe \frac{\omega_n}{T_H} = (2n+1)\pi i + \ln 3
 + \frac{1-i}{\sqrt{n+1/2}} \ \frac{\ell(\ell +1) -1}{18\sqrt 2 \pi^{3/2}}
\ \Gamma^4 (1/4) + \mathcal{O} (1/n) \sioee
which is in agreement with the results from 
a WKB analysis \cite{siob-vdB}
as well as numerical analysis \cite{siob-Nollert}.

%\newpage

\subsection{Kerr black holes}

%\section{Introduction}

Extending the above discussion to rotating (Kerr) black holes
is not straightforward.
Bohr's correspondence principle
\siobe \delta M = \hbar\Re\omega\sioee
and the first law of black hole mechanics
\siobe \delta M = T_H \delta S_{BH} + \Omega\delta J\sioee
imply the asymptotic expression\cite{siob-Hod}
\siobe\label{sioe-1a} \Re\omega = T_H \ln 3 + m\Omega \sioee
where $m$ is the azimuthal eigenvalue of the wave and
$\Omega$ is the angular velocity of horizon.
In deriving the above, we identified $\delta S_{BH} \equiv \ln 3$ \cite{siob-BM}.
Even though the above result has the correct limit as $\Omega\to 0$ (in agreement with the Schwarzschild expression (\ref{sioe-1})), it is in conflict
with numerical results
\cite{siob-BCKO}
indicating $\Re\omega \approx m\Omega$.
%\newpage

To resolve the above contradiction, we shall obtain an analytic solution to the wave (Teukolsky \cite{siob-Teukolsky})
equation
which will be
valid for asymptotic modes bounded from above by $1/a$, where
\siobe a= \frac{J}{M}\sioee 
with $J$ being the angular momentum and $M$ the mass of the Kerr black hole.
The calculation will be valid for
$a\ll 1$ which
includes the Schwarzschild case ($a=0$) \cite{siob-MS2}.
Our results
will confirm Hod's
expression (\ref{sioe-1a})
and not necessarily contradict numerical results
(the latter may still be valid in the asymptotic regime $1/a \lesssim \omega$).
In the Schwarzschild limit ($a\to 0$)
the range of frequencies extends to infinity
and our expression
reduces to the expected form (\ref{sioe-1}).
%\newpage

The metric of a Kerr black hole is
\siobea ds^2 = &-& \left( 1- \frac{2Mr}{\Sigma} \right) dt^2 + \frac{4Mar\sin^2\theta}{\Sigma}\ dtd\phi + \frac{\Sigma}{\Delta}\ dr^2
\nonumber\\
&+& \Sigma d\theta^2 + \sin^2\theta \left( r^2+a^2 + \frac{2Ma^2 r\sin^2\theta}{\Sigma} \right)\
d\phi^2\sioeea
where
$ \Sigma = r^2 + a^2 \cos^2\theta$, $\Delta = r^2 -2Mr + a^2 =(r-r_-)(r-r_+)$
and we have set
Newton's constant $G=1$.
The angular velocity of the horizon and Hawking temperature, respectively, are
\siobe \Omega = \frac{a}{2Mr_+}\ \ , \ \ \ \
T_H = \frac{1 - r_-/r_+}{8\pi M}
\sioee
%\newpage

\subsubsection{Massless perturbations}

Massless perturbations are governed by the Teukolsky wave equation \cite{siob-Teukolsky}
\siobea\left( \frac{(r^2+a^2)^2}{\Delta} - a^2 \sin^2\theta \right) \
\frac{\partial^2\Psi}{\partial t^2} + \frac{4Mar}{\Delta}\ \frac{\partial^2\Psi}{\partial t\partial\phi}
+ \left( \frac{a^2}{\Delta} - \frac{1}{\sin^2\theta}\right) \ \frac{\partial^2\Psi}{\partial\phi^2} & & \nonumber\\
- \frac{1}{\Delta^s} \frac{\partial}{\partial r} \left( \Delta^{s+1}
\frac{\partial\Psi}{\partial r} \right)
 - 2s \left( \frac{M(r^2-a^2)}{\Delta} - r -ia\cos\theta \right)\ \frac{\partial\Psi}{\partial t}
& & \nonumber\\
- \frac{1}{\sin\theta} \frac{\partial}{\partial\theta}
\left( \sin\theta \frac{\partial\Psi}{\partial\theta} \right)
-2s \left( \frac{a(r-M)}{\Delta} + \frac{i\cos\theta}{\sin^2\theta} \right)\
\frac{\partial\Psi}{\partial\phi}
+ (s^2\cot^2\theta - s)\Psi &=& 0\nonumber\\ \sioeea
where $s=0,-1,-2$ for scalar, electromagnetic and gravitational perturbations, respectively.
%We shall keep the discussion general.
%In fact, it will be necessary to avoid integer values of $j$ throughout
%the discussion and only take the limit $s\to$ integer at the end of the
%calculation.
a
Writing the wavefunction in the form
\siobe \Psi = e^{-i\omega t} e^{im\phi} S(\theta) f(r) \sioee
we obtain the
angular equation
$$ \frac{1}{\sin\theta} (\sin\theta\ S')'
 +
\left( a^2\omega^2 \cos^2\theta - \frac{m^2}{\sin^2\theta} -2a\omega s\cos\theta
- \frac{2ms\cos\theta}{\sin^2\theta}
 - s^2 \cot^2\theta \right) S $$
\siobe = -(A+s)S\sioee
where $A$ is the separation constant (eigenvalue) and the
radial equation
\siobe \frac{1}{\Delta^s} (\Delta^{s+1} f')' + V(r) f = (A+a^2\omega^2) f\sioee
where the potential is given by
\siobe V(r) = \frac{(r^2+a^2)^2\omega^2 - 4aMr\omega m + a^2m^2
 +2ia(r-M) ms -2iM (r^2 - a^2) \omega s}{\Delta}
+ 2ir\omega s \sioee
%and $A$ is an eigenvalue to be determined by solving the angular eq.~(\ref{eqang}).
%Before attempting to solve these equations, let us 
Let us simplify the notation by placing the horizon at $r=1$,
i.e., by setting
% This is accomplished by setting
\siobe 2M = 1+a^2 \ \ , \ \ \ \
r_- = a^2 \ \ , \ \ r_+ = 1 \sioee
and solve the two wave equations by expanding in $a$.
We shall keep terms up to $\mathcal{O}(a)$
assuming $\omega$ is large but bounded from above by $1/a$,
($1\lesssim \omega \lesssim 1/a$).
Thus
$\omega$ is in an intermediate
range
which becomes asymptotic in the Schwarzschild limit $a\to 0$.

%Our calculation is valid in the limit of small $a$ ($a\ll 1$).
The solutions to the angular equation to lowest order are spin-weighted spherical
harmonics with eigenvalue
% and the separation constant (eigenvalue) $A$ is
\siobe A = \ell(\ell+1) - s(s+1) + \mathcal{O} (a\omega)\sioee
%It is convenient to express the radial eq.~(\ref{eq5ar}) in terms of a ``tortoise coordinate.''
Near the horizon ($r\to 1$),
\siobe f(r) \sim (r-1)^\lambda \ \ , \ \ \lambda = i(\omega - am) + \mathcal{O} (1/\omega)\sioee
At infinity ($r\to\infty$),
$f(r) \sim e^{i\omega r}$.
Introducing the ``tortoise coordinate''
\siobe z = \omega r + (\omega - am) \ln (r-1) \sioee
the boundary conditions
read
\siobe f(z) \sim e^{\pm iz} \ \ , \ \ z\to \pm\infty
\sioee
%\newpage
From the boundary condition at the horizon we deduce the monodromy for
the function
$\mathcal{F} (z) \equiv e^{iz} f(z)$ (notice that $\mathcal{F} \sim$ const.~as
$z\to +\infty$)
around the singular point
$r=1$,
\siobe \mathcal{M} (1) = e^{4\pi (\omega - am)} + \mathcal{O} (a^2) \sioee
To express the radial equation in terms of the tortoise coordinate, define
\siobe f(r) = \Delta_0^{-s/2}\ \frac{R(r)}{\sqrt{r(\omega r-am)}} \sioee
$\Delta_0 = r(r-1)$ (note $\Delta = \Delta_0 + \mathcal{O} (a^2)$).
Inverting $z=z(r)$,
\siobe r = \sqrt{-\frac{2z}{\omega}} + \mathcal{O} (1/\omega) \sioee
the radial equation
to lowest order in $1/\sqrt\omega$ in terms of $R$ reads
\siobe \frac{d^2R}{dz^2} + \left\{ 1 + \frac{3is}{2z} + \frac{4-s^2
- 4iams}{16z^2} \right\}\ R = 0\sioee
to be solved along the entire real axis.
This is Whittaker's equation. The
solutions may be written as
\siobe M_{\kappa,\pm\mu}(x) = e^{-x/2} x^{\pm\mu + 1/2} M(\siohalf \pm \mu - \kappa, 1\pm 2\mu, x)\sioee
where $\kappa = \frac{3s}{4}$, $\mu^2 = \frac{s(s+4iam)}{16}$,
$M_{\kappa,\pm\mu}$ is Kummer's function (also called $\Phi$) and we set $x= 2iz$.
%\newpage
We need to introduce Whittaker's function
\siobe
W_{\kappa,\mu}(x) =
\frac{\Gamma(-2\mu)}{\Gamma(\siohalf - \mu-\kappa)}\ M_{\kappa,\mu}(x)
+ \frac{\Gamma(2\mu)}{\Gamma(\siohalf +\mu-\kappa)}\ M_{\kappa,-\mu}(x)
\sioee
due to its clean asymptotic behavior,
\siobe W_{\kappa,\mu}(x) \sim e^{-x/2}\ x^\kappa\ (1 + \mathcal{O} (1/x))
\ \ , \ \ \ \  |x|\to\infty~.\sioee
%\newpage
We may compute the monodromy by deforming
the contour as before.
Going around an arc of angle $3\pi$, we have
\siobe M_{\kappa,\pm\mu}(e^{3\pi i} x)
= -i e^{\pm 3\pi i\mu} M_{-\kappa, \pm\mu} (x)\sioee
where we used
$M(a,b,-x) = e^{-x} M(b-a,b,x)$,
therefore
\siobe
W_{\kappa,\mu}(e^{3\pi i} x) =
-i e^{3\pi i \mu}\ \frac{\Gamma(-2\mu)}{\Gamma(\siohalf - \mu-\kappa)}\ M_{-\kappa,\mu}(x)
-i e^{-3\pi i \mu}\ \frac{\Gamma(2\mu)}{\Gamma(\siohalf +\mu-\kappa)}\ M_{-\kappa,-\mu}(x)
\sioee
To find the asymptotic behavior, we need
\siobe M_{-\kappa,\mu} (x) = \frac{\Gamma(1+2\mu)}{\Gamma(\siohalf +\mu +\kappa)}\
e^{-i\pi\kappa} W_{\kappa,\mu} (e^{i\pi} x) +
\frac{\Gamma(1+2\mu)}{\Gamma(\siohalf +\mu -\kappa)}\
e^{-i\pi(\siohalf+\mu +\kappa)} W_{-\kappa,\mu} (x)
\sioee
As $|x|\to\infty$, we obtain
%{\small\siobe M_{-\kappa,\mu} (x) \sim \frac{\Gamma(1+2\mu)}{\Gamma(\siohalf +\mu +\kappa)}\
%e^{-i\pi\kappa}\ e^{x/2}\ (-x)^\kappa +
%\frac{\Gamma(1+2\mu)}{\Gamma(\siohalf +\mu -\kappa)}\
%e^{-i\pi(\siohalf+\mu +\kappa)} e^{-x/2}\ x^{-\kappa}
%\sioee}
%therefore
\siobe W_{\kappa,\mu}(e^{3\pi i} x) \sim A e^{x/2} x^\kappa
+ B e^{-x/2} x^{-\kappa} \sioee
where
\siobe A =
-i e^{3\pi i \mu}\ \frac{\Gamma(-2\mu)}{\Gamma(\siohalf - \mu-\kappa)}\
\frac{\Gamma(1+2\mu)}{\Gamma(\siohalf +\mu +\kappa)}\ e^{-\pi i \kappa}
+ (\mu\to -\mu)
\sioee
and $B$ is not needed for our purposes.
%\siobe \mathcal{B} =
%-i e^{3\pi i \mu}\ \frac{\Gamma(-2\mu)}{\Gamma(\siohalf - \mu-\kappa)}\
%\frac{\Gamma(1+2\mu)}{\Gamma(\siohalf +\mu -\kappa)}\
%e^{-i\pi(\siohalf+\mu +\kappa)}
%+ (\mu\to -\mu)
%\sioee
%\newpage
After some algebra, we deduce
\siobe A = - (1+2\cos \pi s) + \mathcal{O} (a^2)\sioee
where we used the identities
$\Gamma (1-x)\Gamma(x) = \frac{\pi}{\sin\pi x}$,
$\Gamma(\siohalf+x)\Gamma(\siohalf -x) = \frac{\pi}{\cos\pi x}$.
%\siobi
%\item correct Schwarzschild limit
%\item no $\mathcal{O} (a)$ corrections.
%\sioei
The monodromy around $r=1$ is
\siobe \mathcal{M} (1) = e^{4\pi(\omega -ma)} = A
\sioee
therefore \cite{siob-MS2}
\siobe \Re\omega = \frac{1}{4\pi}\ \ln (1+2\cos \pi s) + ma + \mathcal{O} (a^2) \sioee
%\rightline{\small\sl [Musiri, Siopsis]}
%
in agreement with Hod's formula
for gravitational waves ($s=-2$)
in the
small-$a$ limit
(in which $\Omega \approx a$, $T_H \approx \frac{1}{4\pi}$).
However, it should be emphasized that these are not asymptotic values of
QNMs but bounded from above by $1/a$.

\subsubsection{Massive perturbations}

The case of massive perturbations is interesting because it reveals instabilities.
As is well-known, the Schwarzschild spacetime is stable against all kinds of
perturbations, massive or massless
which makes the
%\siobi\item[$\blacktriangleright$]
Schwarzschild geometry appropriate to study astrophysical
objects.
%\sioei
On the other hand, Kerr spacetime
is stable against massless perturbations
but not against massive bosonic fields \cite{siob-Det}. The
instability timescale is much larger than the age of
the Universe so the problem is not expected to have observable consequencies.
Nevertheless, the study of instabilities is an important subject and QNMs provide an indispensable tool.

%\rightline{\small \sl [Detweiler]}

For a massive scalar of mass $\mu$, the radial wave equation reads
%\begin{eqnarray}
\siobe
\frac{d}{ d r}\left(\Delta\frac{dR}{ d r}\right)
+
\left\{
\frac{\omega^2(r^2+a^2)^2-4aMm\omega r+m^2a^2}{\Delta}
-\mu^2 r^2-a^2\omega^2-\ell(\ell+1)
\right\}R=0,
\sioee
%\end{eqnarray}
We are interested in solving this equation for a small mass and low frequencies ($\mu, \omega \ll 1/M$) \cite{siob-Det}.

Away from the horizon ($r\gg M$), we may approximate by
\siobe \frac{d^2}{dr^2} (rR) + \left[ -k^2 + \frac{2M\mu^2}{r} -
\frac{\ell(\ell+1)}{r^2} \right] rR = 0 \ \ , \ \ \ \ k^2 = \mu^2 - \omega^2 \sioee
The solution to this equation is given in terms of a confluent hypergeometric function,
\siobe R(r) = (2kr)^\ell e^{-kr} U(\ell+1-M\mu^2/k,2(\ell+1),2kr) \sioee
Near the horizon ($r\ll \ell/|k|$), we may approximate by
\siobe z(z+1) \frac{d}{dz} \left[ z(z+1) \frac{dR}{dz} \right] +
\left[ P^2 - \ell (\ell+1) z(z+1) \right] R =
0 \sioee
where $P = \frac{am-2Mr_+\omega}{r_+-r_-}$, $z = \frac{r-r_+}{r_+-r_-}$.
The solution to this equation is given in terms of a hypergeometric function,
\siobe R(z) = \left( \frac{z}{z+1} \right)^{iP} F (-\ell, \ell+1; 1-2iP; z+1) \sioee
Matching the two expressions in the overlap region ($M\ll r\ll \ell/|k|$),
we obtain the frequencies
\siobe \omega_n \approx \mu + i\gamma_n \ \ , \ \ \ \ n\in\mathbb{N} \sioee
where
\siobe \gamma_n = \mathcal{C}_{\ell n} \mu (\mu M)^{4(\ell+1)} \frac{am}{M-2\mu r_+}
\prod_{j=1}^\ell \left[ j^2 \left( 1 - \frac{a^2}{M^2} \right)
+ \left( \frac{am}{M} - 2\mu r_+ \right)^2 \right] \sioee
and $\mathcal{C}_{\ell n} = \frac{2^{2(2\ell+1)}
(2\ell+1+n)!(\ell!)^2}{(\ell+1+n)^{2(\ell+2)} (2\ell+1)^2 n!
((2\ell)!)^4}$.

For $m>0$, we have $\gamma_n > 0$ yielding an instability.
For the fastest growing mode (with $\ell=1$, $m=1$, $n=2$ (2p state)) we have
\siobe \tau = \frac{1}{\gamma} = \frac{24}{a\mu^2 (\mu M)^7} \sioee
which is generally large.

Notice that there is no instability
in the Schwarzschild limit ($a\to 0$)
and for massless perturbations ($\mu \to 0$);
in both cases, $\gamma\to 0$ and therefore the lifetime $\tau\to\infty$.

%\newpage

\subsection{Half-integer spin}

In the case of a perturbation of half-integer spin we need to solve the Teukolsky equation \cite{siob-Kh-Ru}
with potential
\siobe\label{eqpot}
V(r) = f(r)\left( \frac{\ell(\ell+1)}{r^2} + \frac{1}{r^3} \right)
+\frac{2i\omega j}{r} - \frac{3i\omega j}{r^2} + \frac{j^2}{4r^4}
\sioee
where $j$ is the spin of the perturbing field (e.g., $j=1/2$ for Dirac fermion).
We shall set $r_0=1$, for simplicity, so $f(r) = 1 - \frac{1}{r}$.

Expanding around the black hole singularity $z=\omega r_* = 0$,
\siobe\label{eqVexp}
\frac{1}{\omega^2}V(z) = \frac{3ij}{2z} - \frac{4-j^2}{16z^2} + \frac{\mathcal{A}}{\omega^{1/2} z^{3/2}} + \mathcal{O} (1/\omega) \ \ , \ \ \ \
\mathcal{A} = \frac{\ell(\ell+1)+ \frac{1-j^2}{3}}{2\sqrt 2 } \sioee
we obtain the zeroth-order wave equation
\siobe \frac{d^2\Psi}{dz^2} + \left[ 1 - \frac{3ij}{2z} - \frac{4-j^2}{16z^2} \right] \Psi
=0 \sioee
whose solutions are the Whittaker functions
\siobe\label{eqpsipm0} \Psi_\pm^{(0)} (z) = M_{\lambda,\pm\mu} (-2iz) \ \ , \ \ \ \
\lambda = \frac{3j}{4} \ \ , \ \ \mu = \frac{j}{4} \sioee
The calculation of the monodromy as before leads to the modes \cite{siob-Kh-Ru}
\siobe\label{eqsp0} \frac{\omega_n}{T_H} = -(2n+1) \pi i + \ln (1+2\cos\pi j) + \mathcal{O} (1/\sqrt n) \sioee
in agreement with the result for integer spin ( which came from the Regge-Wheeler equation).
For a Dirac fermion, $j=1/2$, so
%\siobe\label{eq0j} \frac{\omega_n}{T_H} = -(2n+1) \pi i  + \mathcal{O} (1/\sqrt n)\ \ , \ \ \ \ j+ \frac{1}{2} \in\mathbb{N} \sioee
asymptotically, the real part vanishes.
%\newpage

The first-order correction may also be calculated as before \cite{siob-MS3}.
The result is
\siobea\label{eqsp1} \frac{\omega_n}{T_H} &=& -(2n+1) \pi i + \ln (1+2\cos\pi j) \nonumber\\
&& -\frac{2i}{\sqrt{-in/2}}\ \sin 4\pi\mu\ \frac{\bar b_+ A_- B_- + \bar b_- A_+ B_+}{e^{-4\pi i \mu}A_+ B_- - e^{4\pi i \mu}A_- B_+} + \mathcal{O} (1/n) \sioeea
where
\siobe \bar b_\pm = \frac{\mathcal{A}}{4\mu} \int_0^\infty \frac{dz}{{z}^{3/2}}
M_{\lambda,\pm\mu} (-2iz) M_{\lambda,\pm\mu} (-2iz) \sioee
and $A_\pm = \frac{\Gamma (1\pm 2\mu)}{\Gamma(\frac{1}{2} \pm\mu+\lambda)}
e^{i\pi(\frac{1}{2} \pm\mu-\lambda)}$, $B_\pm = \frac{\Gamma (1\pm
2\mu)}{\Gamma(\frac{1}{2} \pm\mu-\lambda)} e^{-i\pi\lambda}$.
This result appears to be a complicated function of $j$, so let us look at specific cases.

For $j=1/2$ (Dirac fermions), we obtain
\siobe\label{eqsp1aD} \frac{\omega_n}{T_H} = -(2n+1) \pi i +\frac{1+i}{2\sqrt{n}}\ \left( \ell + \frac{1}{2} \right)^2\ \Gamma^2\left( \frac{1}{4} \right)+ \mathcal{O} (1/n) \sioee
%
%\newpage
%\begin{center}
%\includegraphics{../2007-lesbos/D1.eps}
%\end{center}
%\vspace{-2in}
%{\tiny High overtones ($n\ge 100$) of massless Dirac fermions for $\ell+j = 1$.
%Numerical data by Konoplya.}
%\newpage
%{\tiny High overtones ($n\ge 100$) of massless Dirac fermions for $\ell+j = 2$.
%Numerical data by Konoplya.}
%
%\begin{center}
%\includegraphics{../2007-lesbos/D2.eps}
%\vspace{-25in}
%{\tiny High overtones ($n\ge 100$) of massless Dirac fermions for $\ell+j = 2$.
%Numerical data by Konoplya.}
%\end{center}
which is in good agreement with numerical data \cite{siob-MS3}.

%\newpage

For $j=3/2$, we find
\siobe \frac{\omega_n}{T_H} = -(2n+1) \pi i + \mathcal{O} (1/n) \sioee
so there are no first-order corrections to the spectrum.

For $j=5/2$, we have
\siobe \frac{\omega_n}{T_H} = -(2n+1) \pi i
+\frac{1+i}{\sqrt{2n}}\ \mathcal{A} \ \Gamma^2\left( \frac{1}{4} \right)+
\mathcal{O} (1/n) \sioee
etc.

All of the above spectra agree with the general expression we obtained for integer spin using the Regge-Wheeler equation.
The relation of the latter to the Teukolsky equation is worth exploring further.

\section{Anti-de Sitter spacetime}
\label{siosec:3}

%\section{Anti-de Sitter spacetime}
%
%\vspace{0.25in}

According to the AdS/CFT correspondence,
QNMs of AdS black holes are expected to
correspond to perturbations of the dual Conformal Field Theory (CFT) on the boundary.
The
establishment of
such a correspondence is hindered by difficulties in solving the wave equation
governing the various types of perturbation.
In three dimensions one obtains a hypergeometric equation which leads to
explicit analytic expressions for the QNMs \cite{siob-CL,siob-BSS}.
In five dimensions one obtains a Heun equation and a derivation of analytic expressions for QNMs is no longer possible.
On the other hand, numerical results exist in four, five and seven dimensions
\cite{siob-HH,siob-Star,siob-Kono}.

%\newpage
%%%%%%%%%%%%%%%
% page 3

%%%%%%%%%%%%%%%

\subsection{Scalar perturbations}

%\newpage
%\section{Introduction}

To find the asymptotic form of QNMs, we need to find an approximation to the wave equation valid in the
high frequency regime.
In three dimensions the resulting wave equation will be an exact equation (hypergeometric equation).
In five dimensions, we shall turn the Heun equation into a hypergeometric equation
which will lead to
an analytic expression for the asymptotic form of QNM frequencies
in agreement
with numerical results.

%\section{A large AdS black hole}
%\newpage
%\section{Three dimensions}
\subsubsection{AdS$_3$}

In three dimensions the wave equation for a massless scalar field is
\siobe
\frac{1}{R^2\; r}\partial_r \left( r^3 \left( 1- \frac{r_0^2}{r^2}\right) \partial_r \Phi\right) -\frac{R^2}{r^2 - r_0^2 }\partial_t^2 \Phi + \frac{1}{r^2}\partial_x^2 \Phi =0
\sioee
Writing the wavefunction in the form
\siobe
\Phi = e^{i(\omega t-px) }\Psi (y) ,\ \ \ \ \ y = \frac{r_0^2}{r^2}
\sioee
the wave function becomes
\siobe
y^2 (y-1)\left( (y-1) \Psi' \right)'
+ \hat\omega^2\, y\Psi +\hat p^2\, y(y-1)\Psi
=0
\sioee
to be solved in the interval $0<y<1$, where
\siobe
\hat\omega = \frac{\omega R^2}{2r_0} = \frac{\omega}{4\pi T_H},\ \ \ \hat p = \frac{pR}{2r_0} = \frac{p}{4\pi R T_H}\;.
\sioee
%\newpage
For QNMs, we are interested in the
solution
\siobe \Psi (y) = y(1-y)^{i\hat\omega} {}_2F_1 (1+i(\hat\omega + \hat p), 1+i(\hat\omega - \hat p); 2; y)\sioee
which vanishes at the boundary ($y\to 0$).
%\newpage
Near the horizon ($y\to 1$), we obtain a mixture of ingoing and outgoing waves,
%\newline
%{\small\sl [$\because$ standard Hypergeometric function identities]}
\siobe \Psi \sim A_+ (1-y)^{-i\hat\omega} + A_-(1-y)^{+i\hat\omega}\ \ , \ \ \ \
A_\pm =
\frac{\Gamma(\pm 2i\hat\omega)}{\Gamma(1\pm i(\hat\omega + \hat p))\Gamma(1\pm i(\hat\omega - \hat p))}\nonumber\sioee
%\nonumber\\
%B &=& \frac{\Gamma(-2i\hat\omega)}{\Gamma(1-i(\hat\omega + \hat p))\Gamma(1-i(\hat\omega - \hat p))}\nonumber\sioeea
%$\Psi$ linear combination of $\Psi_+$ and $\Psi_-\therefore$
%\siobe \Psi = A\Psi_- + B\Psi_+\sioee
%on account of eq.~(\ref{eq11}).
Setting $A_- =0$, we deduce the quasi-normal frequencies
%For QNMs: $\Psi$ purely ingoing at horizon, so
%set
%\siobe B=0\sioee
%with $B$ given in~(\ref{eq15}).
%Solutions (QNM frequencies):
\siobe \hat\omega = \pm \hat p  -in\quad,\quad n=1,2,\dots \sioee
which form a discrete spectrum of complex frequencies with $\Im\hat\omega < 0$.

%NB: we obtained two sets of frequencies, with opposite $\Re\hat\omega$.

%\newpage
%\section{A large AdS black hole}

\subsubsection{AdS$_5$}

Restricting attention to the case of a large black hole, the massless scalar wave equation reads
\siobe
\frac{1}{r^3}\partial_r (r^5\, f(r)\, \partial_r \Phi) -\frac{R^4}{ r^2\, f(r) }\partial_{t}^2\Phi - \frac{R^2}{r^2}\; \vec\nabla^2\Phi = 0
\ \ , \ \ \ \ \ f(r) = 1- \frac{r_0^4}{r^4}
\sioee
Writing the solution in the form
\siobe 
\Phi = e^{i(\omega t - \vec p\cdot \vec x)} \Psi (y)
\ \ , \ \ \ \
y = \frac{r^2}{r_0^2} 
\sioee
the radial wave equation becomes
\siobe
(y^2-1)\left( y(y^2-1) \Psi' \right)' + \left(\frac{\hat\omega^2}{4}\, y^2 - \frac{\hat p^2}{4}\, (y^2-1)\right)\Psi = 0
\sioee
%Two solutions by examining behavior near the
%horizon ($y\to 1$),
%Near the horizon we obtain in- and out-going waves
%\siobe \Psi_\pm \sim (y-1)^{\pm i\hat\omega/4}\sioee
%where $\Psi_+$ is outgoing and $\Psi_-$ is ingoing.
%Different set by studying
%behavior at large $r$\newline
%($y\to \infty$)
%\siobe \Psi\sim y^{h_\pm} \ \ , \ \ h_\pm = 0,-2\sioee
%so one of
%the solutions contains logarithms.
%
For QNMs, we are interested
in the analytic solution which vanishes at the boundary and behaves as an ingoing wave at the horizon.
The wave equation contains an additional
%$$\Psi\sim y^{-2} \ \ \mathrm{as} \ \ y\to\infty$$
% By considering the other 
(unphysical) singularity at $y=-1$, at which the wavefunction behaves as
%$\Rightarrow$
%another set of solutions
$\Psi \sim (y+1)^{\pm \hat\omega /4}$.
% \ \ \mathrm{near} \ \  y=-1\sioee
%. Following the discussion in the three-dimensional case,
%we shall isolate the behavior at the two singularities $y=\pm 1$ and
Isolating the behavior of the wavefunction near the singularities $y=\pm 1$,
\siobe
\Psi (y) = (y-1)^{-i\hat\omega/4} (y+1)^{\pm\hat\omega/4} F_\pm (y)
\sioee
we shall obtain two sets of modes with the same $\Im\hat\omega$, but opposite
$\Re\hat\omega$.
%, as we shall see (similarly to the $d=3$ case (eq.~(\ref{eqwme2}))).

%It is easily deduced from eqs.~(\ref{eq24}) and (\ref{eq25}) that
$F_\pm (y)$ satisfies the Heun equation
\siobea
y(y^2-1) F\pm'' + \left\{ \left( 3- \frac{i\pm 1}{2}\, \hat\omega \right) y^2 - \frac{i \pm 1}{2}\, \hat\omega y -1 \right\} F_\pm' & & \nonumber\\
+ \left\{ \frac{\hat\omega}{2}\left( \pm \frac{i\hat\omega}{4} \mp 1-i\right) y - (i\mp 1)\frac{\hat\omega}{4} - \frac{\hat p^2}{4} \right\}\; F_\pm &=& 0 
\sioeea
to be solved in a region in the complex $y$-plane
containing $|y|\ge 1$ which
includes the physical regime $r> r_h$.

For large $\hat\omega$, the constant terms in the polynomial coefficients of $F'$ and $F$ are small compared with the other terms,
therefore they may be dropped.
The wave equation may then be approximated by a hypergeometric equation
\siobe
(y^2-1) F_\pm'' + \left\{ \left( 3- \frac{i\pm 1}{2}\, \hat\omega \right) y - \frac{i \pm 1}{2}\, \hat\omega \right\} F_\pm'
+ \frac{\hat\omega}{2}\left( \pm \frac{i\hat\omega}{4} \mp 1-i\right)\; F_\pm =0
\sioee
in the asymptotic limit of large frequencies $\hat\omega$.
The acceptable solution is
\siobe F_0(x) = {}_2F_1 ( a_+, a_-; c; (y+1)/2)
\ \ , \ \ \ a_\pm = 1-{\textstyle{\frac{i \pm 1}{4}}}\,\hat\omega\pm 1
\quad,\quad c = {\textstyle{\frac{3}{2}}}\pm {\textstyle{\frac{1}{2}}}\,\hat\omega\sioee
For proper behavior at the boundary ($y\to\infty$), we demand that $F$ be a {\em polynomial},
which leads to the condition
\siobe a_+ = -n \ \ , \ \ n = 1,2,\dots\sioee
Indeed, it implies that $F$ is a polynomial of order $n$, so as $y\to\infty$,
%at infinity it behaves as 
$F\sim y^n
\sim y^{-a_+}$
%The behavior of $\Psi$ is then deduced from (\ref{eq25}) to be
and $\Psi \sim y^{-i\hat\omega/4} y^{\pm\hat\omega/4} y^{-a_+} \sim y^{-2}$,
as expected.

We deduce the quasi-normal frequencies \cite{siob-MS4}
\siobe \hat\omega = \frac{\omega}{4\pi T_H} = 2n(\pm 1-i) \sioee
in agreement with numerical results.

%{\bf Monodromy argument}

%These frequencies may also be obtained by a monodromy argument similar to the
%one in $d=3$. 
It is perhaps worth mentioning that these frequencies may also be deduced by a simple monodromy argument \cite{siob-MS4}.
Considering the monodromies around the singularities,
if the wavefunction has no singularities other than $y=\pm 1$,
the contour around $y=+1$ may be unobstructedly deformed into the contour
around $y=-1$, which yields
\siobe \mathcal{M} (1) \mathcal{M} (-1) = 1\sioee
Since the respective monodromies are
$\mathcal{M} (1) = e^{\pi \hat\omega /2}$ and $\mathcal{M} (-1) = e^{\mp i\pi \hat\omega /2}$,
using $\Im\hat\omega < 0$, we deduce
$\hat\omega = 2n(\pm 1-i)$, in agreement with our result above.

%\lhead{\fancyplain{}{\bfseries\tiny QNMs - massive modes}}
\subsection{Gravitational perturbations}

%Here we present a fairly comprehensive study of quasi-normal modes of
Next we consider gravitational perturbations of AdS Schwarzschild black holes of arbitrary size in $d$ dimensions.
We shall derive analytic expressions for the asymptotic spectrum \cite{siob-NS} including first-order corrections \cite{siob-MNS}.
Our results will be in good agreement with numerical results.
%AdS Schwarzschild black holes with metric in $d$ dimensions

The metric is
\siobe\label{line}
ds^2 = -f(r)dt^2 +\frac{ dr^2 }{ f(r) } +r^2d \Omega_{d-2}^2\ \ ,\ \ \ f(r) = \frac{r^2}{R^2}+1-\frac{2\mu}{r^{d-3}} \;.
\sioee
%$\blacktriangleright$ derive analytical expressions including
%first-order corrections.
%
%$\blacktriangleright$  results in good agreement with results of
%numerical analysis.

%In this section we discuss gravitational perturbations.
%For massless perturbations, the method discussed in section~\ref{sect1} is
%not directly applicable.
%Instead, we extend the procedure of~\cite{CNS,NS} to include first-order
%corrections to analytical expressions for quasi-normal frequencies.
%Our results are in good agreement with numerical results~\cite{CKL}.

The radial wave equation
% for gravitational perturbations in the black-hole
%background~(\ref{line}) 
can be cast into a Schr\"odinger-like form,
\siobe\label{sch}
  -\frac{d^2\Psi}{dr_*^2}+V[r(r_*)]\Psi =\omega^2\Psi \;,
\sioee
in terms of the tortoise coordinate defined by
\siobe\label{tortoise}
  \frac{dr_*}{dr} = \frac{1}{f(r)}\;.
\sioee
The potential $V$ for the various types of perturbation has been found by Ishibashi and Kodama \cite{siob-IK}.
For tensor, vector and scalar perturbations, we obtain, respectively,
%\rightline{\small \sl [Nat\'ario and Schiappa]}
%{\small
\siobe\label{eqVT} V_{\mathsf{T}} (r) = f(r) \left\{ \frac{\ell (\ell +d-3)}{r^2} + \frac{(d-2)(d-4) f(r)}{4r^2} + \frac{(d-2) f'(r)}{2r} \right\} \sioee
\siobe\label{eqVV} V_{\mathsf{V}}(r) = f(r) \left\{ \frac{\ell (\ell +d-3)}{r^2} + \frac{(d-2)(d-4) f(r)}{4r^2} - \frac{r f'''(r)}{2(d-3)} \right\} \sioee
\siobea\label{eqVS} V_{\mathsf{S}}(r) &=& \frac{f(r)}{4r^2} \left[ \ell (\ell +d-3) - (d-2) + \frac{(d-1)(d-2)\mu}{r^{d-3}} \right]^{-2} \nonumber\\
&\times& \Bigg\{ \frac{d(d-1)^2(d-2)^3 \mu^2}{R^2r^{2d-8}}
- \frac{6(d-1)(d-2)^2(d-4)[\ell (\ell+d-3) - (d-2)] \mu}{R^2r^{d-5}}\nonumber\\
&& + \frac{(d-4)(d-6)[\ell (\ell+d-3) - (d-2)]^2 r^2}{R^2} +
\frac{2(d-1)^2(d-2)^4 \mu^3}{r^{3d-9}}\nonumber\\
&& + \frac{4(d-1)(d-2)(2d^2-11d+18)[\ell (\ell+d-3) - (d-2)]\mu^2}{r^{2d-6}}\nonumber\\
&& + \frac{(d-1)^2(d-2)^2(d-4)(d-6)\mu^2}{r^{2d-6}}
- \frac{6(d-2)(d-6)[\ell (\ell+d-3) - (d-2)]^2 \mu}{r^{d-3}}\nonumber\\
&& - \frac{6(d-1)(d-2)^2(d-4)[\ell (\ell+d-3) - (d-2)] \mu}{r^{d-3}}\nonumber\\
&& + 4 [\ell (\ell+d-3) - (d-2)]^3 + d(d-2) [\ell (\ell+d-3) - (d-2)]^2 \Bigg\} \nonumber\sioeea
%}
%Evidently, the potential always vanishes at the horizon ($V(r_H) = 0$, since $f(r_H)=0$) regardless of the type of perturbation.
Near the black hole singularity ($r\sim 0$),
% the tortoise coordinate~(\ref{tortoise}) may be expanded as
%\be\label{x0}
%r_* = -\frac{1}{(d-2)}\frac{r^{d-2}}{2\mu} - \frac{1}{(2d-5)}\frac{r^{2d-5}}{(2\mu)^2} +\dots
%\ee
%where we have kept the second term in the expansion of $r$ and have chosen the
%integration constant so that $r_*=0$ at $r=0$.
%Using~(\ref{x0}), we may expand the potential near the black hole singularity in the three different cases
%(eqs.~(\ref{eqVT}), (\ref{eqVV}) and (\ref{eqVS})), respectively as
%{\small
\siobe\label{eqVT0} V_{\mathsf{T}} = -\frac{1}{4r_*^2}+\frac{\mathcal{A}_{\mathsf{T}} }{[-2(d-2)\mu]^{\frac{1}{d-2}}} r_*^{-\frac{d-1}{d-2}} + \dots \, , \ \ \ \
  \mathcal{A}_{\mathsf{T}} = \frac{(d-3)^2}{2(2d-5)}+\frac{\ell(\ell+d-3)}{d-2},
\sioee
\siobe\label{eqVV0} V_{\mathsf{V}} = \frac{3}{4r_*^2}+\frac{\mathcal{A}_{\mathsf{V}} }{[-2(d-2)\mu]^{\frac{1}{d-2}}} r_*^{-\frac{d-1}{d-2}} + \dots \ \ , \ \ \ \
\mathcal{A}_{\mathsf{V}} = \frac{d^2-8d+13}{2(2d-15)} + \frac{\ell (\ell +d-3)}{d-2}\sioee
and
\siobe\label{eqVS0}
  V_{\mathsf{S}} =  -\frac{1}{4r_*^2}+\frac{\mathcal{A}_{\mathsf{S}} }{[-2(d-2)\mu]^{\frac{1}{d-2}}} r_*^{-\frac{d-1}{d-2}} + \dots \, ,
\sioee
where
\siobe\label{eqVS0a}
  \mathcal{A}_{\mathsf{S}} = \frac{ (2d^3-24d^2+94d-116)}{4(2d-5)(d-2)}+\frac{ (d^2-7d+14)[ \ell(\ell+d-3)-(d-2)]}{(d-1)(d-2)^2}
\sioee
%}
We have included only the terms which contribute to the order we are interested in.
We may summarize the behavior of the potential near the origin by
\siobe\label{eqV0} V= \frac{j^2 -1}{4r_*^2}+\mathcal{A}\, r_*^{-\frac{d-1}{d-2}} + \dots \sioee
where $j=0$ ($2$) for scalar and tensor (vector) perturbations.
%and the constant coefficient $\mathcal{A}$ can be found from eqs.(\ref{eqVT0}),
%(\ref{eqVV0}), (\ref{eqVS0}) and (\ref{eqVS0a}) in the various cases.
%Throughout the calculation, we shall pretend that $j$ is not an integer.
%At the end of the calculation, we shall let $j\to 0,2$, as appropriate.

On the other hand, near the boundary (large $r$),
\siobe V = \frac{j_\infty^2-1}{4(r_*-\bar r_*)^2} + \dots \ \ , \ \ \ \ \bar r_* = \int_0^\infty \frac{dr}{f(r)} \sioee
where $j_\infty = d-1$, $d-3$ and $d-5$ for tensor, vector and scalar perturbations,
respectively.

After rescaling the tortoise coordinate $(z=\omega r_*)$,
% the Schr\"odinger-like 
the wave equation to first order becomes
%~(\ref{sch}) with the potential~(\ref{eqV0}) becomes
%\be\label{sch-eqn}
%  -\frac{d^2\Psi}{dz^2}+\left[\frac{j^2-1}{4z^2}-1\right]\Psi =-\A
%\; \o^{-\frac{d-3}{d-2}} \, z^{-\frac{d-1}{d-2}}\Psi \, ,
%\ee
%where
%\be\label{A}
%  \A=\frac{\A_{S,T}\;\o^{-\frac{d-3}{d-2}}}{[(d-2)(-2\mu)]^{\frac{1}{d-2}}}.
%\ee
%In the large frequency limit, we may treat the right-hand side of (\ref{sch-eqn}) as a correction.  This will allow us to to solve the equation perturbatively.  We may re-express (\ref{sch-eqn}) as
\siobe\label{we-h}
  \left( \mathcal{H}_0+\omega^{-\frac{d-3}{d-2}} \, \mathcal{H}_1 \right) \Psi =0,
\sioee
where
\siobe\label{H0-H1}
  \mathcal{H}_0= \frac{d^2}{dz^2}-\left[\frac{j^2-1}{4z^2}-1\right]\ \ ,\ \ \mathcal{H}_1=-\mathcal{A}
\; z^{-\frac{d-1}{d-2}}.
\sioee
By treating $\mathcal{H}_1$ as a perturbation, we may expand the wave function
\siobe\label{expandwf}
  \Psi(z)=\Psi_0(z)+\omega^{-\frac{d-3}{d-2}} \, \Psi_1(z)+\dots
\sioee
and solve the wave equation perturbatively.

The zeroth-order wave equation,
\siobe\label{we-0}
  \mathcal{H}_0\Psi_0(z)=0,
\sioee
may be solved in terms of Bessel functions,
\siobe\label{soln_0}
  \Psi_0(z)=A_1\sqrt{z}\, J_{\frac{j}{2}}(z)+A_2 \sqrt{z}\, N_{\frac{j}{2}}(z).
\sioee
For large $z$, it behaves as
%Next, we extend our solution near the horizon to $r\sim\infty\ (z\sim\infty)$.
% we may consider the following asymptotic expansion of  Bessel functions
%\be\label{J-asym}
%  J_{\frac{j}{2}}(y)\sim\sqrt{\frac{2}{\pi z}}\cos(y\mp \a_+)\ \ ,\ \ N_{\frac{j}{2}}(y)\sim\sqrt{\frac{2}{\pi z}}\sin(y\mp \a_+)\ \ ,\ \ y\gg \pm1,
%\ee
%where $\a_\pm = \frac{\pi}{4}(1\pm j)$.
%In this asymptotic limit, the solution (\ref{soln_0}) may be approximated by
\siobea \label{soln-0-0}
  \Psi_0(z)&\sim&  \sqrt{\frac{2}{\pi}}\left[A_1\cos(z-\alpha_+)+A_2\sin(z-\alpha_+)\right]\nonumber\\
  &=&\frac{1}{\sqrt{2\pi}}(A_1-iA_2)e^{-i\alpha_+}e^{iz} + \frac{1}{\sqrt{2\pi}}(A_1+iA_2)e^{+i\alpha_+}e^{-iz}\nonumber
\sioeea
where $\alpha_\pm = \frac{\pi}{4}(1\pm j)$.

At the boundary ($r\to\infty$),
the wavefunction ought to vanish,
therefore the acceptable solution is
\siobe\label{soln-0-infty}
  \Psi_0(r_*) = B\sqrt{\omega(r_*-\bar r_*)}\; J_{\frac{j_\infty}{2}}(\omega (r_*-\bar r_*))
\sioee
Indeed, $\Psi \to 0$ as $r_*\to \bar r_*$, as desired.

Asymptotically (large $z$), it behaves as
\siobe\label{eq54} \Psi(r_*) \sim \sqrt{\frac{2}{\pi}}\, B\cos\left[ \omega(r_*-\bar r_*)+\beta\right] \ , \ \ \ \
\beta =\frac{\pi}{4}(1+ j_\infty) \sioee
We ought to match this to the asymptotic form of the wavefunction in the vicinity of the black-hole singularity along the Stokes line $\Im z = \Im (\omega r_*) = 0$.
This leads to a
constraint on the coefficients $A_1,\ A_2$,
\siobe \label{constraint_1}
  A_1\tan(\omega \bar r_* -\beta -\alpha_+)-A_2=0.
\sioee
By imposing the boundary condition at the horizon
\siobe  \Psi(z) \sim e^{iz}\ \ , \ \ \ \ z\to -\infty\label{bc-0}\ ,
\sioee
we obtain a second constraint.
To find it,
we need to analytically continue the wavefunction near the
black hole singularity ($z=0$) to negative values of $z$.
A rotation of $z$ by $-\pi$ corresponds to a rotation by $-\frac{\pi}{d-2}$ near the origin in the complex $r$-plane.
Using the known behavior of Bessel functions
\siobe\label{eqBrot} J_\nu (e^{-i\pi} z) = e^{-i\pi\nu} J_\nu (z) \ , \ \ \ \
N_\nu (e^{-i\pi} z) = e^{i\pi\nu} N_\nu (z) - 2i\cos \pi\nu\, J_\nu (z)\sioee
for $z<0$ the wavefunction changes to
\siobe\label{soln_0r}
  \Psi_0(z)= e^{-i\pi(j+1)/2} \sqrt{-z}\, \left\{ \left[ A_1 -i (1+e^{i\pi j}) A_2 \right]\, J_{\frac{j}{2}}(-z)+A_2 e^{i\pi j} \, N_{\frac{j}{2}}(-z) \right\}
\sioee
whose asymptotic behavior is given by
\siobe \Psi \sim \frac{e^{-i\pi(j+1)/2}}{\sqrt{2\pi}} \left[ A_1-i(1+2e^{j\pi i}) A_2\right]\, e^{-iz}+\frac{e^{-i\pi(j+1)/2}}{\sqrt{2\pi}} \left[ A_1-iA_2\right]\, e^{iz} \sioee
Therefore we obtain a second constraint
\siobe\label{constraint_2}
  A_1 -i(1+2e^{j\pi i}) A_2 = 0\ \ .
\sioee
The two constraints are compatible provided
\siobe\label{eqcomp}
  \left| \begin{array}{cc}  1 &  -i(1+2e^{j\pi i}) \\
                          \tan(\omega \bar r_*-\beta-\alpha_+) & -1 \end{array}   \right| = 0
\sioee
which yields the quasi-normal frequencies \cite{siob-NS}
\siobe
  \omega \bar r_* =\frac{\pi}{4} (2+j+ j_\infty)-\tan^{-1} \frac{i}{1+2e^{j\pi i}} +n\pi
\sioee
%\rightline{\small\sl [Nat\'{a}rio and Schiappa]}

The first-order correction to the above asymptotic expression may be found by standard perturbation theory \cite{siob-MNS}.
%
%\rightline{\small\sl [Musiri, Ness and Siopsis]}
%
%  =========================================================%
%                               First Order Corrections :Scalar and Tensor
%
%  =======================================================\subsection{First order corrections: Scalar and Tensor}
To first order, the wave equation becomes 
\siobe\label{1stwe1}
  \mathcal{H}_0\Psi_1+\mathcal{H}_1\Psi_0=0
\sioee
%where $\H_0$ and $\H_1$ are given in eq.~(\ref{H0-H1}).
The solution is
%.  The solution to eq.(\ref{1stwe1}) is given by
%{\small
\siobe\label{soln_1}
  \Psi_1(z) = \sqrt{z}\, N_{\frac{j}{2}}(z)\int_0^z dz'\frac{\sqrt{z'}\, J_{\frac{j}{2}}(z')
\mathcal{H}_1\Psi_0(z') }{ \mathcal{W} } -  \sqrt{z}\, J_{\frac{j}{2}}(z)\int_0^z dz'\frac{\sqrt{z'}\, N_{\frac{j}{2}}(z') \mathcal{H}_1\Psi_0(z') }{ \mathcal{W} }
\sioee
%}
where $\mathcal{W} = 2/\pi$ is the Wronskian.
%\[
%  \W = \frac{2}{\pi\o x}.
%\]
%Using~(\ref{soln_0}) and (\ref{soln_1}), we may express the solution to the wave equation (\ref{we-h}) 

The wavefunction to first order reads
%{\small
\siobe\label{soln1st0}
  \Psi(z)=\left\{A_1[1-b(z)] -A_2a_2(z)\right\}\sqrt{z} J_{\frac{j}{2}}(z) +\left\{A_2[1+b(z)]+A_1a_1(z)\right\}\sqrt{z} N_{\frac{j}{2}}(z)
\sioee
%}
where
% the functions $a_1(z)$, $a_2(z)$ and $b(z)$ are given by
%{\small
\siobea
  a_1(z) &=& \frac{\pi\mathcal{A}}{2} \, \omega^{-\frac{d-3}{d-2}}\, \int_0^z dz'\;{z'}^{-\frac{1}{d-2}}J_{\frac{j}{2}}(z') J_{\frac{j}{2}}(z') \nonumber\\
  a_2(z) &=& \frac{\pi\mathcal{A}}{2} \, \omega^{-\frac{d-3}{d-2}}\,  \int_0^z dz'\;{z'}^{-\frac{1}{d-2}}N_{\frac{j}{2}}(z') N_{\frac{j}{2}}(z') \nonumber\\
  b(z) &=& \frac{\pi\mathcal{A}}{2} \, \omega^{-\frac{d-3}{d-2}}\,  \int_0^z dz'\;{z'}^{-\frac{1}{d-2}}J_{\frac{j}{2}}(z') N_{\frac{j}{2}}(z') \nonumber
\sioeea
%}
and $\mathcal{A}$ depends on the type of perturbation.

% Once again, we shall extend $z\rightarrow\infty$ in order to match the solution near $r\sim\infty$.
%In doing so, our coefficients ($a_i,b_1$) are no longer functions of $x$.
%By using the asymptotics of the Bessel functions, the solution 
Asymptotically, it behaves as
\siobe\label{1stsoln}
  \Psi(z)\sim \sqrt{\frac{2}{\pi}}\, [A_1' \cos(z-\alpha_+)+ A_2' \sin(z-\alpha_+)]\ ,
\sioee
where
\siobe\label{eq91} A_1' = [1-\bar b]A_1-\bar a_2 A_2\ \ , \ \ \ \
A_2' = [1+\bar b]A_2+\bar a_1 A_1\sioee
and we introduced the notation
\siobe \bar a_1 = a_1(\infty)\ \ , \ \ \ \ \bar a_2 = a_2(\infty)\ \ , \ \ \ \ \bar b = b(\infty) \ . \sioee
The first constraint is modified to
\siobe\label{newconstr1} A_1' \tan (\omega \bar r_* -\beta -\alpha_+) - A_2' = 0\sioee
Explicitly,
\siobe\label{newconstr1a} [ (1-\bar b)\tan (\omega \bar r_* -\beta -\alpha_+)- \bar a_1 ]A_1 -[1+\bar b +\bar a_2 \tan (\omega \bar r_* -\beta -\alpha_+)]A_2 = 0\sioee
To find the second constraint to first order,
we need to approach the horizon.
This entails a rotation by $-\pi$ in the $z$-plane.
Using
\siobea a_1 (e^{-i\pi} z) &=& e^{-i\pi \frac{d-3}{d-2}} e^{-i\pi j} a_1 (z)\ , \nonumber\\
a_2 (e^{-i\pi} z) &=& e^{-i\pi \frac{d-3}{d-2}} \left[ e^{i\pi j} a_2(z)
- 4 \cos^2 \frac{\pi j}{2} a_1(z) - 2i (1+e^{i\pi j} ) b (z) \right]\ , \nonumber\\
b (e^{-i\pi} z) &=& e^{-i\pi \frac{d-3}{d-2}} \left[ b(z) -i (1+e^{-i\pi j}) a_1(z) \right]\nonumber\sioeea
in the limit $z\to -\infty$ we obtain
\siobe\label{soln1st0r}
\Psi(z) \sim -i e^{-ij\pi/2} B_1 \cos(-z-\alpha_+)
-i e^{ij\pi/2} B_2\sin(-z-\alpha_+)
\sioee
where
%{\small
\siobea
B_1 &=& 
   A_1 -A_1e^{-i\pi\frac{d-3}{d-2}}[{\bar b}-i(1+e^{-i\pi j}){\bar a}_1]
\nonumber \\
& & -A_2e^{-i\pi\frac{d-3}{d-2}} \left[ e^{+i\pi j}{\bar a}_2-4\cos^2\frac{\pi j}{2}{\bar a}_1-2i(1+e^{+i\pi j}){\bar b} \right] \nonumber\\
  & & -i (1+e^{i\pi j})\left[ A_2 +A_2e^{-i\pi\frac{d-3}{d-2}}[{\bar b}-i(1+e^{-i\pi j}){\bar a}_1]+A_1 e^{-i\pi\frac{d-3}{d-2}} e^{-i\pi j}{\bar a}_1 \right] \nonumber\\
B_2 &=& A_2+A_2e^{-i\pi\frac{d-3}{d-2}}[{\bar b}-i(1+e^{-i\pi j}){\bar a}_1]+A_1e^{-i\pi\frac{d-3}{d-2}}e^{-i\pi j}{\bar a}_1\nonumber
\sioeea
%}
Therefore the second constraint to first order reads
%{\small
\siobe\label{newconstr2} [1-e^{-i\pi\frac{d-3}{d-2}}(i
\bar a_1 +\bar b)]A_1 -[i(1+2e^{i\pi j})+e^{-i\pi\frac{d-3}{d-2}}((1+e^{i\pi j})\bar a_1 +e^{i\pi j} \bar a_2-i\bar b)
]A_2 = 0 \sioee
%}
%correcting the zeroth-order constraint~(\ref{constraint_2}).
Compatibility of the two first-order constraints yields
%{\small
\siobe\label{eq99}
  \left| \begin{array}{cc}  1+\bar b+\bar a_2\tan (\omega \bar r_* -\beta -\alpha_+) & i(1+2e^{i\pi j})+e^{-i\pi\frac{d-3}{d-2}}((1+e^{i\pi j})\bar a_1 +e^{i\pi j} \bar a_2-i\bar b) \\
                          (1-\bar b)\tan (\omega \bar r_* -\beta -\alpha_+)- \bar a_1 & 1-e^{-i\pi\frac{d-3}{d-2}}(i
\bar a_1 +\bar b) \end{array}   \right| = 0
\sioee
%}
%\newpage
leading to the first-order expression for quasi-normal frequencies,
\siobea\label{eqo1st}
\omega {\bar r}_* &=& \frac{\pi}{4}(2+j+j_{\infty}) +\frac{1}{2i}\ln 2+n\pi \nonumber\\
   & & -\frac{1}{8}\left\{ 6i\bar b -2i e^{-i\pi\frac{d-3}{d-2}} \bar b  -9\bar a_1+e^{-i\pi\frac{d-3}{d-2}}{\bar a}_1 +{\bar a}_2 - e^{-i\pi\frac{d-3}{d-2}}{\bar a}_2 \right\}
\nonumber\sioeea
where
% we took the limit of interest $j\to 0,2$ wherever it was unambiguous, in
%order to simplify the notation.
%Using
%\be\label{JJ}
%  \int_0^\infty dx\;x^{-\lambda}J_\mu(x)J_\nu(x)=\frac{\G(\lambda)\G(\frac{\nu +\mu+1-\lambda}{2})}{2^\lambda \G(\frac{-\nu +\mu+1+\lambda}{2})\G(\frac{\nu-\mu+1+\lambda}{2})\Gamma(\frac{\nu+\mu+1+\lambda}{2})}\ ,
%\ee
%we obtain explicit expressions for the first-order coefficients,
%{\small
\siobea\label{eqab} \bar a_1 &=& \frac{\pi\mathcal{A}}{4} \left(\frac{n\pi}{2\bar r_*}\right)^{-\frac{d-3}{d-2}} \frac{\Gamma(\frac{1}{d-2})\Gamma(\frac{j}{2}+\frac{d-3}{2(d-2)})}{\Gamma^2(\frac{d-1}{2(d-2)})\Gamma(\frac{j}{2}+\frac{d-1}{2(d-2)})}\nonumber \\
\bar a_2 &=& \left[ 1+2\cot \frac{\pi (d-3)}{2(d-2)} \cot \frac{\pi}{2} \left( -j+\frac{d-3}{d-2}\right) \right]\bar a_1 \nonumber \\
\bar b &=& -\cot \frac{\pi (d-3)}{2(d-2)}\ \bar a_1 \nonumber\sioeea
%}
%where we used the identity $\Gamma (x) \Gamma(1-x) = \frac{\pi}{\sin \pi x}$.
%We also set $\o = n\pi /\bar r_*$, since corrections contribute to higher than first order.
%Notice that these expressions are well-defined when $j$ becomes an integer.
Thus the first-order correction is $\sim \mathcal{O} (n^{-\frac{d-3}{d-2}})$.

%{\Large\boxed{\bf 4d}}
%
The above analytic results are in good agreement with numerical results \cite{siob-CKL}
(see ref.~\cite{siob-MNS} for a detailed comparison).

\subsection{Electromagnetic perturbations}\label{sect3}

The
%The wave equation reduces to~(\ref{sch}) with 
electromagnetic potential in four dimensions is
\siobe\label{V-V}
  V_{\mathsf{EM}} =\frac{\ell(\ell+1)}{r^2}f(r).
\sioee
%where $f(r)$ is given in~(\ref{line}) with $d=4$.
Near the origin,
% this potential may be expanded in terms of the tortoise coordinate. Using eq.~(\ref{x0}), we obtain
\siobe\label{eqVEMr}
  V_{\mathsf{EM}} =\frac{j^2-1}{4r_*^2}+\frac{\ell(\ell+1)r_*^{-3/2}}{2\sqrt{-4\mu}}+\dots \ ,
\sioee
where $j=1$. Therefore we have a vanishing potential to zeroth order.
%Consequently, no analytic expression for quasi-normal frequencies is deduced.
%This is easily seen by substituting $j=1$ in the zeroth-order expression~(\ref{omega-0});
%we obtain a divergent result because $\tan^{-1} i$ is not finite.
%, however, we will leave $j$ general until we evaluate the frequency.  
To calculate the QNM spectrum we need to include first-order corrections from the outset.
%The compatibility condition~(\ref{eq99}) of the two first-order constraints (\ref{newconstr1a}) and
%(\ref{newconstr2}) reads
Working as with gravitational perturbations, we obtain the
QNMs
\siobe\label{eqEMo1}
\omega {\bar r}_* = n\pi -\frac{i}{4}\ln n+\frac{1}{2i}\ln\left( 2(1+i){\cal A}\sqrt{\bar r_*}\right) \ , \ \ \ \ \mathcal{A}
= \frac{\ell(\ell+1)}{2\sqrt{-4\mu}}
\sioee
Notice that the first-order correction behaves as $\ln n$, a fact which may be associated with gauge invariance.

As with gravitational perturbations, the above analytic results are in good agreement with numerical results \cite{siob-CKL}
(see ref.~\cite{siob-MNS} for a detailed comparison).

%\section{Unitarity}
%\label{siosec:4}
%
%\input siounitarity.tex

%\section{Stability}
%\label{siosec:5}
%
%\input siostability.tex

\section{AdS/CFT correspondence and hydrodynamics}
\label{siosec:6}

%\section{Hydrodynamics}
\begin{quotation}
A second unexpected connection comes from studies carried out using the Relativistic Heavy Ion Collider, a particle accelerator at Brookhaven National Laboratory. This machine smashes together nuclei at high energy to produce a hot, strongly interacting plasma. Physicists have found that some of the properties of this plasma are better modeled (via duality) as a tiny black hole in a space with extra dimensions than as the expected clump of elementary particles in the usual four dimensions of spacetime. The prediction here is again not a sharp one, as the string model works much better than expected. String-theory skeptics could take the point of view that it is just a mathematical spinoff. However, one of the repeated lessons of physics is unity - nature uses a small number of principles in diverse ways. And so the quantum gravity that is manifesting itself in dual form at Brookhaven is likely to be the same one that operates everywhere else in the universe.

\rightline{\small -- Joe Polchinski}
\end{quotation}
%\begin{minipage}[b]{.55\linewidth}
%   \includegraphics[angle=0,width=11.8cm]{rhic/rhic.eps}
%  \end{minipage}
%\begin{minipage}[b]{.56\linewidth}
%\centerline{\includegraphics[angle=0,width=12.8cm]{../rhic/2.eps}}
%
%\phantom{a}
%  \end{minipage}
%\newpage

%\newcommand{\qn}{\textswab{q}}
%\newcommand{\wn}{\textswab{w}}
%\newcommand{\<}{\langle}
%\renewcommand{\>}{\rangle}
%\renewcommand{\d}{\partial}
%\newcommand{\N}{{\cal N}}
%\renewcommand{\O}{\hat{\cal O}}
%\newcommand{\q}{\bm{q}}
%\newcommand{\x}{\bm{x}}
%\renewcommand{\Re}{\mathrm{Re}\,}
%\renewcommand{\Im}{\mathrm{Im}\,}
%\newcommand{\gYM}{g_{\mathrm{YM}}}
%\def\ofo{ { {}_2 \! F_1 }}
%\newcommand{\stru}{\rule[-.2in]{0in}{.2in}}

%\subsection{AdS/CFT correspondence and hydrodynamics}

%\rightline{\small\sl [Policastro, Son and Starinets]}

There is a correspondence between
$\mathcal{N}=4$ Super Yang-Mills (SYM) theory in the large $N$ limit
and type-IIB string theory in $\mathrm{AdS}_5\times \mathrm{S^5}$ (AdS/CFT correspondence).
In the low energy limit, string theory is
reduced to classical supergravity and the AdS/CFT correspondence allows one to calculate all
gauge field-theory correlation functions in the strong coupling limit leading to non-trivial predictions on the behavior of gauge theory fluids.
For example,
%$\hookrightarrow$ nontrivial prediction of gauge theory/gravity
%correspondence
%\sioei
the entropy of $\mathcal{N}=4$ SYM
theory in the limit of large
't Hooft coupling is precisely 3/4 its value in the zero
coupling limit.

The long-distance (low-frequency) behavior of any
interacting theory at finite temperature must be described by fluid mechanics 
(hydrodynamics).
This leads to a
universality in physical properties because
hydrodynamics implies very precise constraints on
correlation functions of conserved currents and
the stress-energy tensor.
Their correlators are
fixed once a few transport coefficients are 
known.

\subsection{Hydrodynamics}
\label{sec:hydro}

To study hydrodynamics of the gauge theory fluid, suppose it possesses a conserved current
$j^\mu$.
For simplicity, let us set the
chemical potential $\mu = 0$, so that in thermal equilibrium
the charge density $\langle j^0\rangle =0$.
The retarded thermal Green
function is given by
\siobe
  G^R_{\mu\nu} (\omega, q) = -i\!\int\!d^4x\,e^{-iq\cdot x}\,
  \theta(t) \langle [j_\mu(x),\, j_\nu(0)] \rangle \,,
\sioee
where $q=(\omega,\vec{q})$, $x=(t,\vec{x})$.
It
%$\blacktriangleright$
determines the response of the system to a small
external source coupled to the current.  
For small $\omega$ and $|\vec{q}|$,
the external perturbation varies slowly in space and time.
Then a macroscopic hydrodynamic description for its evolution is possible \cite{siob-PSS}.

For a charged density obeying the diffusion equation
\siobe
  \partial_0 j^0 = D \nabla^2 j^0\,,
\sioee
where $D$ is the diffusion constant with dimension of length,
we obtain an overdamped mode with dispersion relation
\siobe
  \omega = - i D \vec{q}^2 \,,
\sioee
The corresponding retarded Green function has a pole at $\omega=-iD\vec{q}^2$
in the complex $\omega$-plane.

Another important conserved current is the stress-energy tensor
$T^{\mu\nu}$.
Its conservation law may be written as
\siobe
\begin{split}
  &  \partial_0 \tilde T^{00} + \partial_i T^{0i} = 0\,,\\
  &  \partial_0 T^{0i} + \partial_j \tilde T^{ij} = 0\,,
\end{split}
\sioee
where
\siobe
\begin{split}
  \tilde T^{00} &= T^{00} - \rho, \qquad \rho =\langle T^{00}\rangle \,,\\
  \tilde T^{ij} &= T^{ij} - p\delta^{ij}
%= \\ &
   =- \frac1{\rho + p}\Bigl[\eta \Bigl(\partial_i T^{0j} + \partial_j T^{0i}
    -\frac23
    \delta^{ij}\partial_k T^{0k}\Bigl) 
% \\&\qquad\qquad\qquad
+ \zeta \delta^{ij}\partial_k T^{0k}\Bigr],
\end{split}
\sioee
and $\rho$ ($p$) is the energy density (pressure) of the fluid,
$\eta$ ($\zeta$) is its shear (bulk) viscosity. 

One obtains two types of eigenmodes,
the
shear modes which consist of transverse fluctuations of the momentum
density $T^{0i}$, with a purely imaginary eigenvalue
\siobe
  \omega = - i Dq^2 \ \ , \ \ \ \ D = \frac{\eta}{\rho+p} \,,
\sioee
and a sound wave due to simultaneous fluctuations of
the energy density $T^{00}$ and the longitudinal component of momentum
density $T^{0i}$, with dispersion relation
\siobe
  \omega = u_s q - \frac i2 \frac1{\rho+p} 
  \left(\zeta+\frac43\eta\right)q^2 \,,
  \qquad u_s^2 = \frac{\partial p}{\partial\rho}\, .
\sioee
In a conformal field theory,
the stress-energy tensor is traceless, so
\siobe\rho=3p\ \ , \ \ \ \  \zeta=0\ \ , \ \ \ \ u_s = \frac{1}{\sqrt 3}\sioee

%\section{R-charge diffusion}
%\label{sec:Rcharge}

\subsection{Branes}
\label{sec:grav}

To understand the gravitational side of the AdS/CFT correspondence, consider
a non-extremal 3-brane which is a solution of type-IIB low energy equations of motion.
In the near-horizon limit $r\ll R$ where $R$ is the AdS radius, the metric becomes
\siobe
%\begin{split}
ds^2_{10} = 
%{ r_0^2 \over R^2  u}
  \frac{(\pi T R)^2}u
\left( -f(u) dt^2 + dx^2 + dy^2 +dz^2 \right) 
%\\&\quad
 +\frac{R^2}{ 4 u^2 f(u)} du^2 + R^2 d\Omega_5^2\,,
%\end{split}
\sioee
where $T = \frac{r_0}{\pi R^2}$ is the Hawking temperature, and we have defined
$u = \frac{r_0^2}{r^2}$, $f(u)=1-u^2$. 
The horizon corresponds to $u=1$ whereas spatial infinity is at $u=0$.

According to the gauge theory/gravity correspondence,
the above background metric with non-extremality parameter $r_0$
is dual to ${\cal N}=4$ $SU(N)$ 
SYM at finite temperature  $T$
in the limit of $N\rightarrow \infty$, $g^2_{YM}N \rightarrow \infty$.
For the retarded Green function
\siobe
  G_{\mu\nu,\lambda\rho} (\omega, \vec{q})
  = -i\!\int\!d^4x\,e^{-iq\cdot x}\,
  \theta(t) \langle [T_{\mu\nu}(x),\, T_{\lambda\rho}(0)] \rangle\,.
\sioee
we deduce by considering an appropriate perturbation of the background metric \cite{siob-PSS},
%Then, according to the prescription given in~\cite{gamma_paper}, the
%(retarded) two-point function 
%of the stress-energy tensor (in Fourier space) reads
%\begin{equation} 
%  G_{xy,xy}(\omega,\q) =
%  - {\pi^2 N^2  T^4\over 4 } 
%  \lim_{u\rightarrow 0} {f\over u} 
%  \phi_k^* \phi_k' = 
%  - {i \pi N^2 \omega  T^3 \over 8 }  \,.
%\end{equation}
\siobe
  G_{xy,xy}(\omega,\vec{q}) =
  - \frac{ N^2 T^2 }{ 16 }\left( i\, 2 \pi T \omega + \vec{q}^2 \right)  \,.
\sioee
leading to the shear viscosity of strongly coupled 
${\cal N}=4$ SYM plasma (Kubo formula)
\siobe
\eta = \lim_{\omega \rightarrow 0} \frac{1}{ 2\omega} 
 \int\!dt\,d\vec{x}\, e^{i\omega t}\,
\langle [ T_{xy}(x),\,  T_{xy}(0)]\rangle =  
  \frac{\pi}8 N^2 T^3\,.
\sioee
%$\blacktriangleright$ Deduce
Other correlators may also be found by different perturbations of the metric.
One obtains
\begin{subequations}\label{corr-T}
\begin{eqnarray}
  G_{tx,tx}(\omega,\vec{q}) &=&  
  \frac{ N^2\pi T^3 \vec{q}^2 }{ 8 (i\omega - {\cal D} \vec{q}^2)}
+ \dots
\,,
\rule[-.2in]{0in}{.2in}\nonumber\\
 G_{tx,xz}(\omega,\vec{q})  &=& 
- \frac{ N^2\pi T^3 \omega |\vec{q}| }{ 8 (i\omega - {\cal D} \vec{q}^2)}
+ \dots
\,,\rule[-.2in]{0in}{.2in}\nonumber\\
 G_{xz,xz}(\omega,\vec{q}) &=& 
 \frac{ N^2\pi T^3 \omega^2  }{ 8 (i\omega - {\cal D} \vec{q}^2)}
+ \dots
\,,
\end{eqnarray}
\end{subequations}
where
%the formulas are valid up to corrections of order 
%${\cal O} (\textswab{w}^2, \textswab{w}\textswab{q}^2, \textswab{q}^4 )$, and
${\cal D} = \frac1{4\pi T}$.

%{\color{Green} $\blacktriangleright$ half} the diffusion constant for the R-charge.

From the above results, one may deduce the viscosity $\eta$.
Indeed,
recall from hydrodynamics
${\cal D}=\frac{\eta}{\rho+p}$.
Using the
entropy
\siobe
  s = \frac34 s_0 = \frac{\pi^2}2 N^2 T^3\,,
\sioee
where $s_0$ is the entropy at zero coupling,
and the thermodynamic equations
$s=\frac{\partial P}{\partial T}$,
$\rho=3p$, we deduce
$\rho+p=\frac{\pi^2}2N^2T^4$,
therefore
\siobe\eta=\frac\pi8N^2T^3\ \ , \ \ \ \ \frac{\eta}{s} = \frac{1}{4\pi}\sioee
which agrees
with the Kubo formula.
It should be pointed out that there is no agreement unless $s= \frac{3}{4} s_0$, a fact which is still poorly understood.

The above result for the viscosity is based on the gravity dual of the gauge theory fluid and should correspond to its strong coupling regime.
At weak coupling, one obtains by a direct calculation
\siobe \frac{\eta}{s} \gg \frac{1}{4\pi} \sioee
Thus the viscocity coefficient $\eta$ varies
as a function of the 't Hooft coupling,
\siobe
  \eta = f_\eta(g_{\mathrm{YM}}^2N) N^2T^3
\sioee
where $f_\eta(x)\sim \frac{1}{-x^2\ln x}$ for $x\ll1$ and
$f_\eta(x)=\frac\pi8$ for $x\gg1$.

\subsection{Schwarzschild black holes}

%\subsubsection{Conformal soliton flow}

In the metric considered above, the horizon was flat. This corresponds to the limit of a large black hole.
For a black hole of finite size, the horizon is generally a sphere.
Then the boundary of spacetime is $S^3\times\mathbb{R}$.
This may be conformally mapped onto a flat Minkowski space.
Then by holographic renormalization,
the AdS$_5$-Schwarzschild black hole is dual to a spherical shell of plasma on the four-dimensional Minkowski space which first contracts and then expands (conformal soliton flow) \cite{siob-Princ}.

Quasi-normal modes govern the properties of this plasma with long-lived modes (i.e., of small $\Im\omega$) having the most influence.
For example, one obtains the ratio
\siobe \frac{v_2}{\delta} = \frac{1}{6\pi} \Re \frac{\omega^4-40\omega^2+72}{\omega^3-4\omega} \sin \frac{\pi\omega}{2} \sioee
where
$v_2 = \langle \cos 2\phi \rangle$ evaluated at $\theta = \frac{\pi}{2}$ (mid-rapidity) and averaged with respect to the energy density at late times;
$\delta = \frac{\langle y^2 - x^2 \rangle}{\langle y^2 + x^2 \rangle}$ is the eccentricity at time $t=0$.
Numerically, $\frac{v_2}{\delta} = 0.37$, which compares well with the result from RHIC data, $\frac{v_2}{\delta} \approx 0.323$ \cite{siob-PHENIX}.

Another observable is the thermalization time which is found to be
\siobe \tau = \frac{1}{2|\Im \omega|} \approx \frac{1}{8.6 T_{\mathrm{peak}}} \approx 0.08~\mathrm{fm/c}\ \ , \ \ \ \ T_{\mathrm{peak}} = 300~\mathrm{MeV} \sioee
not in agreement with the RHIC result $\tau \sim 0.6$~fm/c \cite{siob-RHIC}, but still encouragingly small.
For comparison, the corresponding result from perturbative QCD is $\tau \gtrsim 2.5$~fm/c \cite{siob-QCD,siob-QCD2}.

The above results motivate the calculation of low-lying QNMs. Earlier, we calculated analytically the asymptotic form of QNMs for large black holes.
We obtained frequencies which were proportional to the horizon radius $r_0$.
We found an infinite spectrum, however we missed the lowest frequencies which
are inversely proportional to $r_0$.
The latter are important in the understanding of the hydrodynamic behavior of the gauge theory fluid via the AdS/CFT correspondence.

\subsubsection{Vector perturbations}

We start with vector perturbations and work in a $d$-dimensional Schwarzschild background. It is convenient to introduce
the coordinate \cite{siob-ego}
\siobe\label{eqru} u = \left( \frac{r_0}{r} \right)^{d-3} \sioee
The wave equation becomes
\siobe\label{eq13} -(d-3)^2 u^{\frac{d-4}{d-3}}\hat f(u) \left( u^{\frac{d-4}{d-3}}\hat f(u) \Psi' \right)' +\hat V_{\mathsf{V}} (u)\Psi = \hat\omega^2 \Psi  \ \ , \ \ \ \
\hat\omega = \frac{\omega}{r_0}\sioee
where prime denotes differentiation with respect to $u$ and we have defined
%{\small
\siobe\label{eq14} \hat f(u) \equiv \frac{f(r)}{r^2} = 1- u^{\frac{2}{d-3}} \left( u- \frac{1 - u}{r_0^2} \right) \sioee
\siobe\label{eq15} \hat V_{\mathsf{V}} (u) \equiv \frac{V_{\mathsf{V}}}{r_0^2} = \hat f(u) \left\{ \hat L^2 + \frac{(d-2)(d-4)}{4} u^{-\frac{2}{d-3}}\hat f(u) - \frac{(d-1)(d-2)\left( 1+ \frac{1}{r_0^2} \right)}{2} u\right\} \sioee
%}
where
$\hat L^2 = \frac{\ell (\ell +d-3)}{r_0^2} $.

%For simplicity,
First let us consider the large black hole limit $r_0 \to\infty$ keeping $\hat\omega$ and $\hat L$ fixed~(small).
%The wave equation simplifies to
%\[ - (d-3)^2 (u^{\frac{2d-8}{d-3}} -u^3)\Psi'' - (d-3) [ (d-4)u^{\frac{d-5}{d-3}}-(2d-5)u^2]\Psi' \]
%\be + \left\{ \hat L^2 + \frac{(d-2)(d-4)}{4}u^{-\frac{2}{d-3}} - \frac{3(d-2)^2}{4} u - \frac{\hat\omega^2}{1-u^{\frac{d-1}{d-3}}} \right\} \Psi = 0 
%\ee
Factoring out the behavior at the horizon ($u=1$)
% (horizon) and $u=0$ (boundary),
\siobe \Psi = (1-u)^{-i \frac{\hat\omega}{d-1}} F(u) \sioee
the wave equation simplifies to
%we obtain
\siobe\label{sch2} \mathcal{A} F'' + \mathcal{B}_{\hat\omega} F' + \mathcal{C}_{\hat\omega , \hat L} F = 0 \sioee
where
%{\small
\siobea \mathcal{A} &=& - (d-3)^2 u^{\frac{2d-8}{d-3}} (1-u^{\frac{d-1}{d-3}}) \nonumber\\
\mathcal{B}_{\hat\omega} &=& - (d-3) [ d-4-(2d-5)u^{\frac{d-1}{d-3}}]u^{\frac{d-5}{d-3}} - 2(d-3)^2 \frac{i\hat\omega}{d-1}\frac{u^{\frac{2d-8}{d-3}} (1-u^{\frac{d-1}{d-3}})}{1-u} \nonumber\\
\mathcal{C}_{\hat\omega , \hat L} &=& \hat L^2 + \frac{(d-2)[d-4-3(d-2)u^{\frac{d-1}{d-3}}]}{4}u^{-\frac{2}{d-3}} \nonumber\\
& & - \frac{\hat\omega^2}{1-u^{\frac{d-1}{d-3}}} + (d-3)^2 \frac{\hat\omega^2}{(d-1)^2}\frac{u^{\frac{2d-8}{d-3}} (1-u^{\frac{d-1}{d-3}})}{(1-u)^2}\nonumber\\
& &
- (d-3)\frac{i\hat\omega}{d-1} \frac{[ d-4-(2d-5)u^{\frac{d-1}{d-3}}]u^{\frac{d-5}{d-3}} }{1-u} - (d-3)^2 \frac{i\hat\omega}{d-1}\frac{u^{\frac{2d-8}{d-3}} (1-u^{\frac{d-1}{d-3}})}{(1-u)^2}\nonumber\sioeea
%}
We may
%For small $\hat\omega$, $\hat L$, eq.~(\ref{sch2}) may be solved perturbatively by writing it as
solve this equation perturbatively by separating
\siobe (\mathcal{H}_0 + \mathcal{H}_1) F = 0 \sioee
where
\siobea\label{eqH0} \mathcal{H}_0 F &\equiv& \mathcal{A} F'' + \mathcal{B}_0 F' + \mathcal{C}_{0 , 0} F \nonumber\\
\mathcal{H}_1 F &\equiv& (\mathcal{B}_{\hat\omega} - \mathcal{B}_0) F' + (\mathcal{C}_{\hat\omega , \hat L} - \mathcal{C}_{0 , 0}) F \nonumber\sioeea
Expanding the wavefunction perturbatively,
\siobe F = F_0 + F_1 + \dots \sioee
at zeroth order the wave equation reads
\siobe\label{eq22} \mathcal{H}_0 F_0 = 0 \sioee
whose acceptable solution is
\siobe\label{eq23} F_0 = u^{\frac{d-2}{2(d-3)}} \sioee
being regular at both the horizon ($u=1$) and the boundary ($u=0$, or $\Psi \sim r^{-\frac{d-2}{2}}\to 0$ as $r\to\infty$).
The Wronskian is
\siobe \mathcal{W} = \frac{1}{u^{\frac{d-4}{d-3}} (1-u^{\frac{d-1}{d-3}})} \sioee
and another linearly independent solution is
\siobe\label{eqchF0} \check F_0 = F_0\int \frac{\mathcal{W}}{F_0^2} \sioee
which is unacceptable because it diverges at both the horizon ($\check F_0 \sim \ln (1-u)$ for $u\approx 1$) and the boundary ($\check F_0 \sim u^{-\frac{d-4}{2(d-3)}}$ for $u\approx 0$, or $\Psi \sim r^{\frac{d-4}{2}} \to\infty$ as $r\to\infty$).
%It may be expressed in terms of hypergeometric functions but its explicit form is not needed for our purposes (first-order perturbation theory).

At first order the wave equation reads
\siobe \mathcal{H}_0 F_1 =- \mathcal{H}_1 F_0 \sioee
whose solution may be written as
\siobe\label{eqF1} F_1 = F_0\int \frac{\mathcal{W}}{F_0^2} \int \frac{F_0\mathcal{H}_1 F_0}{\mathcal{A}\mathcal{W}} \sioee
The limits of the inner integral may be adjusted at will
because this amounts to adding an arbitrary amount of the unacceptable solution.
To ensure regularity at the horizon, choose one of the limits of integration at $u=1$
rendering the integrand regular at the horizon.
Then at the boundary ($u=0$),
\siobe F_1 = \check F_0 \int_0^1 \frac{F_0\mathcal{H}_1 F_0}{\mathcal{A}\mathcal{W}} + \mathrm{regular~terms} \sioee
%where we omitted regular terms.
The coefficient of the singularity ought to vanish,
\siobe\label{eq29} \int_0^1 \frac{F_0 \mathcal{H}_1 F_0}{\mathcal{A}\mathcal{W}} = 0 \sioee
which yields a constraint on the parameters (dispersion relation)
\siobe\label{eqdisp} \mathbf{a}_0 \hat L^2 -i \mathbf{a}_1 \hat\omega - \mathbf{a}_2 \hat\omega^2 = 0 \sioee
After some algebra, we arrive at
\siobe\label{eqcoef} \mathbf{a}_0 = \frac{d-3}{d-1} \ \ , \ \ \ \
\mathbf{a}_1 = d-3 \sioee
The coefficient $\mathbf{a}_2$
may also be found explicitly for each dimension $d$,
but it cannot be written as a function of $d$ in closed form.
It does not contribute to the dispersion relation at lowest order.
E.g., for $d=4,5$, we obtain, respectively
\siobe\label{eqa2} \mathbf{a}_2 = \frac{65}{108} -\frac{1}{3}\ln 3 \ \ , \ \ \ \
\frac{5}{6}-\frac{1}{2}\ln 2 \sioee
%\sioei
%\[ \mathbf{a}_0 = \int_0^1 u^{\frac{2}{d-3}} = \frac{d-3}{d-1}\]
%\[ \mathbf{a}_1 = -(d-3)\int_0^1 u^{\frac{2}{d-3}} \left[ (d-2) \frac{u^{\frac{2d-8}{d-3}} -u^3}{u(1-u)} + \frac{ (d-4)u^{\frac{d-5}{d-3}}-(2d-5)u^2}{1-u}
% + (d-3) \frac{u^{\frac{2d-8}{d-3}} -u^3}{(1-u)^2}\right] \]
%\[ = - (d-1)(d-3) \]
%\[ \mathbf{a}_2 = \int_0^1 u^{\frac{2}{d-3}}\left[ (d-3)^2 \frac{u^{\frac{2d-8}{d-3}} -u^3}{(1-u)^2} - \frac{(d-1)^2}{1-u^{\frac{d-1}{d-3}}} \right] \]
%\[ - (1-u^{d-1}) F'' + \left\{ (d-1) u^{d-2} -2 i \frac{\hat\omega}{d-1} \frac{1-u^{d-1}}{1-u} \right\} F' \]
%For $d=5$, $\mathbf{a}_2 = \frac{5}{6}-\frac{1}{2}\ln 2$.
%For $d=4$, $\mathbf{a}_2 = \frac{65}{108} -\frac{1}{3}\ln 3$.
Eq.~(\ref{eqdisp}) is quadratic in $\hat\omega$ and has two solutions,
\siobe \hat\omega_0 \approx -i\frac{\hat L^2}{d-1} \ \ , \ \ \ \  \hat\omega_1 \approx -i \frac{d-3}{\mathbf{a}_2} + i\frac{\hat L^2}{d-1} \sioee
%where we omitted terms of higher order in $\hat L^2$.
In terms of the frequency $\omega$ and the quantum number $\ell$,
\siobe\label{eq34} \omega_0 \approx -i\frac{\ell(\ell+d-3)}{(d-1)r_0} \ \ , \ \ \ \  \frac{\omega_1}{r_0} \approx -i \frac{d-3}{\mathbf{a}_2} + i\frac{\ell(\ell+d-3)}{(d-1)r_0^2} \sioee
The smaller of the two, $\omega_0$,
is inversely proportional to the radius of the horizon
and is not included in the asymptotic spectrum.
The other solution, $\omega_1$,
is a crude estimate of the first overtone in the asymptotic spectrum, nevertheless
%
%Numerically, for $d=5$,
%\be\label{eqod5} \frac{\omega_1}{r_0} \approx -4.109 i + i \frac{\ell (\ell +2)}{4r_0^2} \ee
%to be compared with the numerical value (\ref{eq2}).
%
%It should be noted that the crude estimate (\ref{eqod5}) already 
it shares two important features with the asymptotic spectrum:
it is proportional to $r_0$
and its dependence on $\ell$ is $\mathcal{O} (1/r_0^2)$.
The approximation may be improved by including higher-order terms.
This increases the degree of the polynomial in the dispersion relation (\ref{eqdisp}) whose roots then yield approximate values of more QNMs.
This method reproduces the asymptotic spectrum derived earlier albeit not in an efficient way.

To include finite size effects,
we shall use perturbation theory (assuming $1/r_0$ is small) and replace
$\mathcal{H}_1$ by
\siobe\label{eqH1} \mathcal{H}_1' = \mathcal{H}_1 + \frac{1}{r_0^2} \mathcal{H}_H \sioee
where
%For finite $r_+$, we may add
\siobe \mathcal{H}_H F \equiv \mathcal{A}_H F'' + \mathcal{B}_H F' + \mathcal{C}_H F \sioee
The coefficients may be easily deduced by collecting $\mathcal{O} (1/r_0^2)$ terms in the exact wave equation.
% given by (\ref{eq13}), (\ref{eq14}) and (\ref{eq15}).
We obtain
\siobea \mathcal{A}_H &=& -2(d-3)^2 u^2(1-u) \nonumber\\
\mathcal{B}_H &=& -(d-3) u\left[ (d-3)(2-3u) - (d-1) \frac{1-u}{1-u^{\frac{d-1}{d-3}}} u^{\frac{d-1}{d-3}} \right] \nonumber\\
\mathcal{C}_H &=& \frac{d-2}{2} \left[ d-4-(2d-5)u - (d-1) \frac{1-u}{1-u^{\frac{d-1}{d-3}}} u^{\frac{d-1}{d-3}} \right] \nonumber\sioeea
Interestingly, the zeroth order wavefunction $F_0$ is an eigenfunction of $\mathcal{H}_H$,
\siobe \mathcal{H}_H F_0 = -(d-2) F_0 \sioee
therefore the first-order finite-size effect is a simple shift of the angular momentum operator
\siobe \hat L^2 \to \hat L^2 - \frac{d-2}{r_0^2} \sioee
The QNMs of lowest frequency are modified to
\siobe\label{eqo0} \omega_0 = - i \frac{\ell(\ell+d-3)-(d-2)}{(d-1)r_0} + \mathcal{O} (1/r_0^2) \sioee
For $d=4, 5$, we have respectively,
\siobe \omega_0 = - i \frac{(\ell-1)(\ell+2)}{3r_0} \ \ , \ \ \ \  - i \frac{(\ell+1)^2-4}{4r_0} \sioee
in agreement with numerical results \cite{siob-CKL,siob-Princ}.

According to the AdS/CFT correspondence,
dual to the AdS Schwarzschild black hole is a gauge theory fluid on the boundary of AdS ($S^{d-2} \times~\mathbb{R}$).
Consider the fluid dynamics ansatz
\siobe\label{eqansv} u_i = \mathcal{K} e^{-i\Omega \tau} \mathbb{V}_i \sioee
where $u_i$ is the (small) velocity of a point in the fluid,
and $\mathbb{V}_i$ a vector harmonic on $S^{d-2}$.
Demanding that this ansatz satisfy the standard equations of linearized hydrodynamics,
one arrives at a constraint on the frequency of the perturbation $\Omega$ which yields \cite{siob-MP}
\siobe \Omega = -i \frac{\ell(\ell+d-3)-(d-2)}{(d-1)r_0} + \mathcal{O} (1/r_0^2) \sioee
in perfect agreement with its dual counterpart.

\subsubsection{Scalar perturbations}

Next we consider scalar perturbations which are calculationally more involved but phenomenologically more important because theirspectrum contains the lowest frequencies.
For a scalar perturbation we ought to replace the potential $\hat V_{\mathsf{V}}$ by
%{\small
\siobea  \hat V_{\mathsf{S}}(u) &=& \frac{\hat f(u)}{4} \left[ \hat m + \left( 1 + \frac{1}{r_0^2} \right) u\right]^{-2}  \nonumber\\
&\times& \Bigg\{ d(d-2) \left( 1+ \frac{1}{r_0^2} \right)^2 u^{\frac{2d-8}{d-3}}
- 6(d-2)(d-4)\hat m \left( 1+ \frac{1}{r_0^2} \right) u^{\frac{d-5}{d-3}}\nonumber\\
&& + (d-4)(d-6)\hat m^2u^{-\frac{2}{d-3}} +
(d-2)^2 \left( 1+ \frac{1}{r_0^2} \right)^3 u^3 \nonumber\\
&& + 2(2d^2-11d+18)\hat m \left( 1+ \frac{1}{r_0^2} \right)^2 u^2\nonumber\\
&& + \frac{(d-4)(d-6)\left( 1+\frac{1}{r_0^2} \right)^2}{r_0^2} u^2
- 3(d-2)(d-6)\hat m^2 \left( 1+\frac{1}{r_0^2} \right) u\nonumber\\
&& - \frac{6(d-2)(d-4)\hat m\left( 1+\frac{1}{r_0^2} \right)}{r_0^2} u + 2 (d-1)(d-2)\hat m^3 + d(d-2) \frac{\hat m^2}{r_0^2} \Bigg\} \nonumber\\ \sioeea
%}
where
$\hat m = 2\frac{\ell (\ell+d-3) - (d-2)}{(d-1)(d-2)r_0^2} = \frac{2(\ell + d-2)(\ell -1)}{(d-1)(d-2)r_0^2} $.

In the large black hole limit $r_0\to \infty$ with $\hat m$ fixed (small), the potential simplifies to
%{\small
\siobea  \hat V_{\mathsf{S}}^{(0)}(u) &=& \frac{1-u^{\frac{d-1}{d-3}}}{4( \hat m + u)^2}
\Bigg\{ d(d-2) u^{\frac{2d-8}{d-3}}
- 6(d-2)(d-4)\hat m u^{\frac{d-5}{d-3}}\nonumber\\
&& + (d-4)(d-6)\hat m^2u^{-\frac{2}{d-3}} +
(d-2)^2 u^3\nonumber\\
&&
+ 2(2d^2-11d+18)\hat m u^2
- 3(d-2)(d-6)\hat m^2  u
+ 2 (d-1)(d-2)\hat m^3  \Bigg\} \nonumber\\ \sioeea
%}
%\siobi\item[$\blacktriangleright$]
The wave equation has an additional singularity due to the double pole of the scalar potential
at $u = -\hat m$.
%\item[$\blacktriangleright$]
It is desirable to factor out the behavior not only at the horizon, but also at the boundary and the pole of the scalar potential,
%We therefore define
\siobe\label{eqPsiF} \Psi = (1-u)^{-i\frac{\hat\omega}{d-1}} \frac{u^{\frac{d-4}{2(d-3)}}}{\hat m + u} F(u) \sioee
%\sioei
Then the wave equation reads
%Factoring out the bahavior at the horizon, we obtain
\siobe\label{eqwsc} \mathcal{A} F'' + \mathcal{B}_{\hat\omega} F' + \mathcal{C}_{\hat\omega} F = 0 \sioee
where
%{\small
\siobea \mathcal{A} &=& - (d-3)^2 u^{\frac{2d-8}{d-3}} (1-u^{\frac{d-1}{d-3}}) \nonumber\\
\mathcal{B}_{\hat\omega} &=& - (d-3) u^{\frac{2d-8}{d-3}} (1-u^{\frac{d-1}{d-3}}) \left[ \frac{d-4}{u} -\frac{2(d-3)}{\hat m + u} \right] \nonumber\\
&& - (d-3) [ d-4-(2d-5)u^{\frac{d-1}{d-3}}]u^{\frac{d-5}{d-3}} - 2(d-3)^2 \frac{i\hat\omega}{d-1}\frac{u^{\frac{2d-8}{d-3}} (1-u^{\frac{d-1}{d-3}})}{1-u} \nonumber\\
\mathcal{C}_{\hat\omega} &=&  - u^{\frac{2d-8}{d-3}} (1-u^{\frac{d-1}{d-3}}) \left[ -\frac{(d-2)(d-4)}{4 u^2} - \frac{(d-3)(d-4)}{u(\hat m + u)} + \frac{2(d-3)^2}{(\hat m + u)^2} \right] \nonumber\\
&& - \left[ \left\{ d-4-(2d-5)u^{\frac{d-1}{d-3}} \right\} u^{\frac{d-5}{d-3}} + 2(d-3) \frac{i\hat\omega}{d-1}\frac{u^{\frac{2d-8}{d-3}} (1-u^{\frac{d-1}{d-3}})}{1-u}\right]\left[ \frac{d-4}{2u} - \frac{d-3}{\hat m + u} \right] \nonumber\\
&& - (d-3)\frac{i\hat\omega}{d-1} \frac{[ d-4-(2d-5)u^{\frac{d-1}{d-3}}]u^{\frac{d-5}{d-3}} }{1-u} - (d-3)^2 \frac{i\hat\omega}{d-1}\frac{u^{\frac{2d-8}{d-3}} (1-u^{\frac{d-1}{d-3}})}{(1-u)^2}\nonumber\\
&& + \frac{\hat V_{\mathsf{S}}^{(0)}(u)-\hat\omega^2}{1-u^{\frac{d-1}{d-3}}} + (d-3)^2 \frac{\hat\omega^2}{(d-1)^2}\frac{u^{\frac{2d-8}{d-3}} (1-u^{\frac{d-1}{d-3}})}{(1-u)^2}\nonumber\sioeea
We shall define zeroth-order wave equation as $\mathcal{H}_0 F_0 = 0$, where
%as in (\ref{eq22}) with
\siobe\label{eq0sc} \mathcal{H}_0 F \equiv \mathcal{A} F'' + \mathcal{B}_0 F' \sioee
The acceptable zeroth-order solution is
\siobe\label{eqF0sc} F_0(u) = 1 \sioee
%\siobi\item[$\blacktriangleright$]
which is plainly regular at all singular points ($u=0,1, -\hat m$).
%\item[$\blacktriangleright$]
It corresponds to a wavefunction vanishing at the boundary
($\Psi \sim r^{-\frac{d-4}{2}}$ as $r\to\infty$).
%\sioei

The Wronskian is
\siobe\label{eqWsc} \mathcal{W} = \frac{\left( \hat m + u\right)^2 }{u^{\frac{2d-8}{d-3}} (1-u^{\frac{d-1}{d-3}})} \sioee
and an nacceptable solution is $ \check F_0 = \int \mathcal{W} $.
It can be written in terms of hypergeometric functions.
For $d\ge 6$, it has a singularity at the boundary, $\check F_0 \sim u^{-\frac{d-5}{d-3}}$ for $u\approx 0$,
or $\Psi \sim r^{\frac{d-6}{2}}\to\infty$ as $r\to\infty$.
For $d=5$, the acceptable wavefunction behaves as $r^{-1/2}$ whereas the unacceptable one behaves as $r^{-1/2} \ln r$.
For $d=4$, the roles of $F_0$ and $\check F_0$ are reversed, however the results still valid because the correct boundary condition at the boundary is a Robin boundary condition \cite{siob-ego,siob-MP}.
Finally, we note that $\check F_0$ is also singular (logarithmically) at the horizon ($u=1$).

Working as in the case of vector modes,
we arrive at the first-order constraint
\siobe\label{eqcnssc} \int_0^1 \frac{\mathcal{C}_{\hat\omega}}{\mathcal{A}\mathcal{W}} = 0 \sioee
%where we used (\ref{eq29}), (\ref{eqF0sc}) and
because $\mathcal{H}_1 F_0 \equiv (\mathcal{B}_{\hat\omega} - \mathcal{B}_0) F_0'
+ \mathcal{C}_{\hat\omega} F_0 = \mathcal{C}_{\hat\omega} $.
This leads to the
dispersion relation
\siobe\label{eqcnssc2} \mathbf{a}_0 - \mathbf{a}_1 i\hat\omega - \mathbf{a}_2 \hat\omega^2 = 0 \sioee
After some algebra, we obtain
\siobe \mathbf{a}_0 = \frac{d-1}{2} \ \frac{ 1+ (d-2)\hat m}{(1+ \hat m )^2} \ \ ,
\ \ \ \ \mathbf{a}_1 = \frac{d-3}{(1+ \hat m )^2} \ \ ,
\ \ \ \ \mathbf{a}_2 = \frac{1}{\hat m} \left\{ 1 + O(\hat m) \right\} \sioee
For small $\hat m$, the quadratic equation has solutions
\siobe\label{eqosc1} \hat\omega_0^\pm \approx - i\frac{d-3}{2} \ \hat m \pm \sqrt{ \frac{d-1}{2} \ \hat m} \sioee
related to each other by $\hat\omega_0^+ = -\hat\omega_0^{-*}$, which is a
general symmetry of the spectrum.

%Neither solution is included in the spectrum obtained from asymptotic QNMs.
%To obtain approximations to those modes, we need to consider higher orders in perturbation theory.
%Notice also that unlike vector modes, these scalar lowest frequency modes have finite real part.
%Using (\ref{eqm}), we may express the frequencies 
%in terms of the quantum number $\ell$,
%\be\label{eqosc1a} \omega_0^\pm \approx - i(d-3) \ \frac{\ell (\ell+d-3)-(d-2)}{(d-1)(d-2) r_0} \pm \sqrt{ \frac{\ell (\ell+d-3)-(d-2)}{d-2}} \ee
%Thus the real part is independent of $r_0$ whereas the imaginary part is inversely proportional to $r_0$ (for $r_0\gtrsim 1$).

Finite size effects
at first order amount to a shift of the coefficient $\mathbf{a}_0$ in the dispersion relation
\siobe \mathbf{a}_0 \to \mathbf{a}_0 + \frac{1}{r_0^2} \mathbf{a}_H \sioee
After some tedious but straightforward algebra, we obtain
\siobe \mathbf{a}_H = \frac{1}{\hat m} \left\{ 1 + O(\hat m) \right\} \sioee
The modified dispersion relation yields the modes
\siobe\label{eqosc2} \hat\omega_0^\pm \approx - i\frac{d-3}{2} \ \hat m \pm \sqrt{ \frac{d-1}{2} \ \hat m +1} \sioee
%correcting (\ref{eqosc1}).
In terms of the quantum number $\ell$,
%Explicitly,
\siobe\label{eqosc2a} \omega_0^\pm \approx - i(d-3) \ \frac{\ell (\ell+d-3)-(d-2)}{(d-1)(d-2) r_0} \pm \sqrt{ \frac{\ell (\ell+d-3)}{d-2}} \sioee
%correcting (\ref{eqosc1a}).
in agreement with numerical results \cite{siob-Princ}.

%\rightline{\small\sl [Friess, Gubser, Michalogiorgakis and Pufu]}

Notice that the imaginary part is inversely proportional to $r_0$, as in vector case.
In the scalar case, we also obtained a
finite real part independent of $r_0$.
It yields the
speed of sound $v_s = \frac{1}{\sqrt{d-2}}$ which is the correct value in the presence of conformal invariance.

Turning to the implications of the above results for the AdS/CFT correspondence,
we may
perturb the gauge theory fluid on the boundary of AdS ($S^{d-2} \times \mathbb{R}$) using the ansatz
\siobe\label{eqanss} u_i = \mathcal{K} e^{-i\Omega \tau} \nabla_i \mathbb{S} \ \ , \ \ \ \
\delta p = \mathcal{K}' e^{-i\Omega \tau} \mathbb{S} \sioee
where $u_i$ is the (small) velocity of a point in the fluid
and $\delta p$ is a pressure perturbation.
They are both given in terms of $\mathbb{S}$, a scalar harmonic on $S^{d-2}$.
Demanding that this ansatz satisfy the equations of linearized hydrodynamics,
one obtains a frequency of perturbation $\Omega$ in perfect agreement with our analytic result \cite{siob-ego,siob-MP}.

\subsubsection{Tensor perturbations}

Finally, for completeness we discuss the case of tensor perturbations.
Unlike the other two cases, the asymptotic spectrum of tensor perturbations is the entire spectrum.
To see this,
% let us change variables to (\ref{eqru}) and go to the large black hole limit.
note that in the large black hole limit, the wave equation reads
%{\small
\siobea\label{eqwavt} - (d-3)^2 (u^{\frac{2d-8}{d-3}} -u^3)\Psi'' - (d-3) [ (d-4)u^{\frac{d-5}{d-3}}-(2d-5)u^2]\Psi' && \nonumber\\
 + \left\{ \hat L^2 + \frac{d(d-2)}{4}u^{-\frac{2}{d-3}} + \frac{(d-2)^2}{4} u - \frac{\hat\omega^2}{1-u^{\frac{d-1}{d-3}}} \right\} \Psi &=& 0 
\nonumber\sioeea
%}
For the zeroth-order equation, we may set $\hat L = 0 = \hat\omega$.
The resulting equation may be solved exactly.
Two linaerly independent solutions are ($\Psi = F_0$ at zeroth order)
%{\small
\siobe F_0(u) = u^{\frac{d-2}{2(d-3)}} \ \ , \ \ \ \ \check F_0(u) = u^{-\frac{d-2}{2(d-3)}} \ln \left( 1-u^{\frac{d-1}{d-3}} \right) \sioee
%}
Neither behaves nicely at both ends ($u=0,1$).
Therefore both are unacceptable which makes it
impossible to build a perturbation theory to calculate small frequencies which are inversely proportional to $r_0$.
This negative result is
in agreement with numerical results \cite{siob-CKL,siob-Princ} and in accordance with the
AdS/CFT correspondence.
Indeed,
there is no ansatz that can be built from tensor spherical harmonics $\mathbb{T}_{ij}$ satisfying the linearized hydrodynamic equations, because of the conservation and tracelessness properties of $\mathbb{T}_{ij}$.

\section{Conclusion}
\label{siosec:7}

We discussed the calculation of analytic asymptotic expressions for quasi-normal modes of various perturbations of black holes in asymptotically flat as well as anti-de Sitter spaces.
We also showed how perturbative corrections to the asymptotic expressions can be systematically calculated.

In view of the AdS/CFT correspondence, in AdS spaces we concentrated on low frequency modes because they govern the hydrodynamic behavior of the gauge theory fluid which is dual to the black hole.
Thus, these modes provide a powerful tool in understanding the hydrodynamics of a gauge theory at strong coupling.
%We are beginning to understand non-perturbative QCD effects using string theory
They may lead to experimental consequences pertaining to the quark-gluon plasma produced in heavy ion collisions at RHIC and the LHC.

%\begin{acknowledgment}
%Research supported in part by the US Department of Energy under grant DE-FG05-91ER40627.
%\end{acknowledgment}

%%%%%%%%%%%%%%%%%%%%%%%% referenc.tex %%%%%%%%%%%%%%%%%%%%%%%%%%%%%%
% sample references
% %
% Use this file as a template for your own input.
%
%%%%%%%%%%%%%%%%%%%%%%%% Springer-Verlag %%%%%%%%%%%%%%%%%%%%%%%%%%
%
% BibTeX users please use
% \bibliographystyle{}
% \bibliography{}
%

%\input{referenc}
\end{document}